\documentclass[a4paper, 12pt]{article}

\usepackage{graphicx}
\usepackage{epstopdf}
\usepackage{amsmath}
\usepackage{subcaption}
\usepackage{caption}
\usepackage[round, sort]{natbib}
\usepackage{color}
\usepackage{bm}
\usepackage{authblk}
\usepackage[bottom]{footmisc}
\usepackage[normalem]{ulem}

\let\oldequation\equation
\let\endoldequation\endequation

\renewenvironment{equation}
  {\begin{footnotesize}\oldequation}
  {\endoldequation\end{footnotesize}}

\title{Deformation and orientation of a capsule with viscosity contrast in linear flows: a theoretical study}
\author{Paul Regazzi\thanks{Email address for correspondence: paul.regazzi@univ-amu.fr}\hspace{5pt} and Marc Leonetti\thanks{Email address for correspondence: marc.leonetti@univ-amu.fr}}
\affil{Aix Marseille Univ, CNRS, CINAM, Marseille, France}

\begin{document}
\maketitle
\begin{abstract}
We develop a perturbation theory to study the shape and the orientation of an initially spherical capsule of radius $R$ with a viscosity contrast, a surface tension $\sigma$ and a bending rigidity $\kappa$ in linear flows. The elastic mechanical response of membrane to deformations is described by three elastic constitutive law which are either Hookean, Neohookean or Skalak type leading to the introduction of a surface shear elastic modulus $G_s$ and the Poisson ratio (or analog quantities). At the leading order, the deformation, i.e. the so-called Taylor parameter is proportional to the elastic capillary number $Ca$ which evaluates the ratio between the external viscous stress and the elastic membrane response. In this linear regime, the results do not depend on the elastic constitutive law as expected. Without surface tension and bending rigidity, we recover the results of \cite{barthes1981time} and notably the fact that the Taylor parameter does not depend on the viscosity contrast $\lambda$ contrary to the case of a viscous droplet. In our more general model, the deformation does no longer depend on $\lambda$ at the upper order. Now, the Taylor parameter also depends on two other dimensionless numbers: the surface elastocapillary ratio $\sigma/G_s$ and the dimensionless bending rigidity $B=\kappa/G_sR^2$. At the further order, the angle of inclination of the capsule with the direction of the shear flow, the analog of the Chaffey and Brenner equation for droplets is determined in each case. The results are in excellent agreement with the numerical ones performed with a code based on the boundary integral method providing an useful method to valid numerical developments.
\end{abstract}
\vfill
\section{Introduction}
Conveying, protecting and delivering active molecules are general goals encountered in many fields leading to the building of new smart multiphasic materials which can be passive or active. These ones often mimic nature. Indeed, migrating cells, organelles or cells carried by living fluids offered some guides to pave the way to such new structures. One of them, the capsule or the microcapsule has attracted much attention from several decades (\cite{bah2020fabrication}) as a specific elastic body  or as a biomimetic model of cells, Red Blood Cells in particular (\cite{yazdani2011phase,misbah2012vesicles,becic2025origin,sui2008dynamic}), but also due to its many analogs in cosmetics, health and food science/design/manufacturing (\cite{sagis2015microencapsulation}) and building materials (\cite{tyagi2011development}). Capsules belong to the large family of coated drops, whose the mechanical properties mainly depend on those of the interface. Other examples are liposomes, vesicles (\cite{has2021vesicle,faizi2022vesicle}), polymersomes (\cite{dionzou2016comparison,discher2006polymersomes}), cells (\cite{lu2025transient,dabagh2020localization}), droploon (\cite{ginot2022peg}) and so on.\\

Capsules are droplets bounded by thin elastic shells (\cite{barthes2016motion,dupre2017one}) that preserve their integrity and their ability to be a (bio-)chemical mini-reactor and may avoid any contamination from outside. More generally, the mechanic response of the shell to deformations and stress depends on its material properties driving by strain-hardening or strain-softening elasticity (\cite{de2015stretching}) and two-dimensional (2D) dissipation characterized by surface viscosities (\cite{de2016tank,gires2016transient}). Some capsules exhibit a viscoplastic behavior (\cite{xie2017interfacial}). Several experimental configurations have been studied: capsules in constrictions (\cite{chu2011comparison,risso2006experimental,rorai2015motion,chen2023robust,le2017squeezing}), capsules in shear flow (\cite{chang1993experimentala}), capsules in extensional flow (\cite{chang1993experimentalb,de2014mechanical}), capsules under compression and capsules under electric field (\cite{karyappa2014deformation}). In some cases, the rupture of the capsule is observed, an essential characteristic linked to the material properties and the applied mechanical tension along the shell (\cite{chachanidze2023breakups,chang1993experimentalb,el2024microrheometric,feng2024viscosity,joung2020synthetic,husmann2005deformation}). In shear and extensional flows, the shell can be under compression leading to buckling or wrinkling instabilities (\cite{rehage2002two,walter2001shear}) if the bending resistance is not too high. Moreover, as shown in other elastocapillary systems (\cite{style2017elastocapillarity,bico2018elastocapillarity}) and wrinkling instability, surface tension can play an essential role in the emergence of such patterns.\\

Pioneering theoretical works have been developed to gain insight in the understanding on how deforms a capsule in linear flows in the limit of weak deformations. They determined the deformation at the leading order, i.e. proportional to the capillary number $Ca$ which compares the viscous hydrodynamic stress to the shear elastic response: \cite{barthes1981time,barthes2002effect}. All these developments have been validated further by numerical simulations and they have been essential to determine the shear elastic modulus by comparison with experiments in the linear regime of deformations.\\

In this paper, a theoretical study is performed on initially spherical capsules under linear flows with additional properties compared to previous studies and up to the second order: viscosity contrast different from 1, bending rigidity and the surface tension. The analysis is done in the limit of weak deformation and in the regime of Stokes flow. The physical quantities are developed according to the capillary number up to the second order to allow the determination of the orientation of the capsule in a shear flow. This is the analog of the Chaffey-Brenner relation for a droplet (\cite{chaffey1967second}) but for a capsule with additional properties to be the most general. Moreover, the Hooke, NeoHookean and Skalak constitutive laws are considered. At the leading order, the expressions of the deformations of \cite{barthes1981time} and \cite{barthes2002effect} are recovered without bending energy and surface tension. In chapter 2, all the basic equations in the bulk and at the interface are recalled. In chapter 3, the main analytic steps are provided in a generalized way. In chapter 4, the results (lengths of the equivalent ellipsoid, the deformations in each plane, the angle of orientation) are calculated and discussed. Our results are in excellent agreement with numerical ones.\\

\section{Problem statement and numerical procedure}
\setcounter{subsection}{-1}
\subsection{Notations}
The summation convention is adopted that is each repeated indices in the equations are summed over. Also, the superscript $*$ is concerning the quantities of the internal fluid. The lack of this superscript is referring to the external fluid.
\subsection{Problem statement}
We consider an initially spherical capsule immersed in a neutrally buoyant Newtonian fluid of viscosity $\eta$ and also filled by a fluid of viscosity $\eta^*$ of the same kind. The position of the capsule membrane is characterized by the position vector $\bm{x}$. Because of the object shape, it becomes natural to use spherical coordinates. Based on the requirement that the capsule deformation is small, its radius is written
\begin{equation}
    r = \sqrt{\bm{x}\cdot\bm{x}} = R\left( 1+F\right)
\end{equation}
with $R$ the radius of the undeformed sphere and $|F|\ll 1$ the deformation\footnote{This quantity is expressed as $J$ in \cite{barthes1981time}} dependent on the sphere angles that will be determined later. Naturally, the vector position should contain the deformation tensor. The radius can be used to determine the expression for the normal vector pointing out of the sphere
\begin{equation}
    \bm{n} = \frac{\bm{\nabla} r}{|\bm{\nabla} r|} = \frac{\bm{x}}{r} -r\bm{\nabla} F
\end{equation}
and the tensor projection onto the surface
\begin{equation}
    \bm{P} = \bm{I}-\bm{n}\otimes \bm{n}^T
\end{equation}
where $\bm{I}$ is the identity matrix.
The outer fluid will undergo a linear flow such as shear flow, extensional flow and so on. Far from the capsule, the velocity field can be expressed as
\begin{equation}
    \bm{v}^\infty = \dot\epsilon \left( \bm{E}+\bm{\Omega}\right)\cdot\bm{x}\text{ for }|\bm{x}|\rightarrow+\infty
\end{equation}
the far-field velocity can be decomposed into tensors $E$ which encodes the symmetric properties of the fluid and $\Omega$ the antisymmetric properties. $\dot\epsilon$ is the flow shear rate. The pressure far from the capsule is denoted $p^\infty$. According to those definitions, one can also define the velocity field $v$ that is the relative difference of the flow velocity according to the far-field velocity. Naturally this field as well as the pressure should fade out in the case of far-field
\begin{equation}
    \lim_{|\bm{x}|\rightarrow\infty}\bm{v} = \bm{0},\qquad \lim_{|\bm{x}|\rightarrow\infty}p = 0
    \end{equation}

Assuming the radius of the capsule is sufficiently small, the velocity and pressure fields obey Stokes equations
\begin{equation}
    \Delta \bm{v} = \eta \bm{\nabla} p\quad\quad\bm{\nabla}\cdot\bm{v} = 0
    \label{eq:StokesExt}
\end{equation}
\begin{equation}
    \Delta \bm{v}^* = \eta^* \bm{\nabla} p^*\quad\quad\bm{\nabla}\cdot\bm{v}^* = 0    \label{eq:StokesInt}
\end{equation}
The couple $(\bm{v}^\infty,\,p^\infty)$ satisfies the Eq.\ref{eq:StokesExt}.\\
From now on the equations will be expressed at the interface of the capsule. To determine the expressions for the velocity fields, one could look at the continuity of the fields leading to the following equation
\begin{equation}
    \bm{v}^\infty+\bm{v} = \bm{v}^*\label{eq:velocity-continuity}
\end{equation}
Under the flow strength, the object studied has to undergo a rate of deformation directly counterbalanced by viscoelastic forces. Assuming $\bm{\Pi}$ is the stress tensor, one can write at the interface
\begin{equation}
    \left(\eta(\bm{\Pi}+\bm{\Pi}^\infty)-\eta^*\bm{\Pi}^*\right)\cdot\bm{n} = \bm{f}\label{eq:stress-continuity}
\end{equation}
for $\bm{\Pi}^\infty$ the stress tensor far from the capsule and $\bm{f}$ the forces exerted on the capsule. The stress tensors are
\begin{equation}
    \bm{\Pi} = -p \bm{I} +2\eta\left(\bm{\nabla} \bm{v}+\bm{\nabla} \bm{v}^T \right)
\end{equation}
\begin{equation}
    \bm{\Pi}^* = -p^* \bm{I} +2\eta^*\left(\bm{\nabla} \bm{v}^*+\bm{\nabla} \bm{v}^{*T} \right)
\end{equation}
\begin{equation}
    \bm{\Pi}^\infty = -p^\infty \bm{I} +2\eta\left(\bm{\nabla} \bm{v}^\infty+\bm{\nabla} \bm{v}^{\infty T} \right)
\end{equation}
Then, the evolution equation at steady state gives
\begin{equation}
    \left(\bm{v}+\bm{v}^\infty\right)\cdot\bm{n} = 0\label{eq:evolution-equation}
\end{equation}
The system of equations given by equations (\ref{eq:velocity-continuity}), (\ref{eq:stress-continuity}) and (\ref{eq:evolution-equation}) is sufficient to describe the whole system. To solve this system, it is necessary to determine the constitutive laws used in the capsule's forces.
\subsection{Two-dimensional elastic constitutive laws}

The elastic response of the membrane depends on its deformations, i.e. the variations between the reference position $\bm{X}$ and the same $\bm{x}$ in the deformed state. One should introduce a new tensor $\bm{K}$ encoding information about the evolution of a displacement on the surface regarding the reference state. For instance, in the reference state the normal vector is written $\bm{N}$.
Then one can write the surface displacement gradient
\begin{equation}
    \bm{A} = \left( \bm{I}-\bm{n} \otimes\bm{n}^T \right)\cdot\frac{\partial \bm{x}}{\partial \bm{X}}\cdot\left( \bm{I} - \bm{N}\otimes\bm{N} ^T\right)
\end{equation}
The right Cauchy-Green tensor is deduced
\begin{equation}
    \bm{C} = \bm{A}^T \cdot\bm{A}
\end{equation}
Finally one obtains the surface Green-Lagrange deformation tensor
\begin{equation}
    \bm{e} = \frac{1}{2}\left[\bm{C}-\left( \bm{I}-\bm{N}\otimes\bm{N}^T\right)\right]
\end{equation}
To establish the constitutive laws, the method of \cite{barthes1980motion}, \cite{barthes2002effect} is used by defining the following Jacobian $J$ and strain invariants $I_1$, $I_2$ as
\begin{equation}
    J = \sqrt{\det{\left( \bm{C}+\bm{N}\otimes\bm{N}^T\right)}},\qquad I_1 = 2\text{tr}{(\bm{e})},\qquad I_2 = J^2-1
\end{equation}
Then \cite{barthes2002effect} defined the Cauchy tensor $T$ with strain energy function $w$ as
\begin{equation}
    \bm{T} = \frac{2}{J}\left\{\frac{\partial w}{\partial I_1}\bm{A}\cdot\bm{A}^T+\frac{\partial w}{\partial I_2}J^2\left(\bm{I}-\bm{n}\otimes\bm{n}^T\right)\right\}
\end{equation}
The strain energy function will directly implement the constitutive laws which here read :
\begin{equation}
    \text{(Skalak)}\qquad w^{SK} = \frac{G^{SK}_s}{4}\left(I_1^2+2I_1-2I_2+CI_2^2\right)
\end{equation}
\begin{equation}
    \text{(Hooke)}\qquad w^{H} = \frac{G^{H}_s}{4}\left(2I_1-2I_2+\frac{1}{1-\nu}I_1^2\right)
\end{equation}
\begin{equation}
    \text{(Neo-Hookean)}\qquad w^{NH} = \frac{G^{NH}_s}{2}\left(I_1-1+\frac{1}{I_2+1}\right)
\end{equation}
where $G^{SK}_s$ and $C$ are the surface material properties in the framework of the Skalak Law. $G^{H}_s$ is the surface shear modulus and $\nu$ the surface Poisson ratio in the Hooke model. $G^{NH}_s$ is the surface shear elastic modulus in the framework of the Neo-Hookean model.\\

Inserting one constitutive law inside the Cauchy tensor and applying the surface derivative one obtains the viscoelastic force
\begin{equation}
    \bm{f_{el}} = \bm{P}\cdot\bm{\nabla}\cdot\bm{T}
\end{equation}
Thus, after simplifications the force is
\begin{equation}
    \bm{f_{el}} = -\frac{2}{J}\left(\frac{\partial w}{\partial I_1}+J^2\frac{\partial w}{\partial I_2}\right)\bm{n}\cdot\bm{\nabla} \bm{n} + \bm{P}\cdot\bm{\nabla} \left(\frac{2}{J}\left[\frac{\partial w}{\partial I_1}+J^2\frac{\partial w}{\partial I_2}\right]\right)+ \bm{P}\cdot\bm{\nabla}\cdot\left[\frac{2}{J}\frac{\partial w}{\partial I_1}\left(\bm{A}\cdot\bm{A}^T-\bm{I}\right)\cdot\bm{P}\right]
    \label{eq:fmembrane}
\end{equation}
One can note that the first term involves the surface mean curvature and is thus the intrinsic force according to the pure geometry of the object. The second term is the tangential contribution from the constitutive law. The third term is the stretch of the membrane along the spatial coordinates.

\subsection{Bending force and capillary force: bending rigidity, spontaneous curvature and surface tension}

Although negligible at macroscopic scales, capillary forces can become essential at small scales, especially if triple lines are present. In the case of some capsules, the presence of triple lines at the surface or inside a porous structure with immiscible internal and external fluids such as oil and water for example (\cite{chachanidze2022structural}) leads to a contribution of surface tension. We simply introduce as usual the following free energy:
\begin{equation}
    F_\sigma = \int_S\,\sigma\,dS
\end{equation}
where $\sigma$ is the surface tension. Note that here, $\sigma$ is not the Lagrange multiplier associated to the condition of local incompressibility as in the model of vesicles. In most cases, the bending forces are negligible compared to the in-plane elastic stresses described in the previous chapter. When membrane domains are under compression, their shapes can be unstable and results from a competition with bending forces. Consequently, it is of interest to quantify the contribution of bending to the steady shape. Here, we simply used the model of Helfrich providing the following free energy as in many studies on capsules (\cite{helfrich1973elastic,zhong1989bending,powers2010dynamics}):
\begin{equation}
    F_\kappa = \frac{\kappa}{2}\int_S\,(2H)^2\,dS\,+\,\kappa_g\,\int\,K_g\,dS
\end{equation}
where $H$ and $K_g$ are the mean and gaussian curvatures of the membrane surface respectively. $\kappa$ and $\kappa_g$ are the elastic bending modulus and the elastic gaussian modulus, respectively. The second term only depends on topology. The membrane forces derived from previous free energies are (\cite{yazdani2012three,sui2007transient}):
\begin{equation}
    \bm{f_\sigma} = -2\sigma\,H\,\bm{n}
    \label{eq:force_tension}
\end{equation}
\begin{equation}
    \bm{f_\kappa}\,=\,2\kappa\,\left(2H^3\,-\,2K_g\,H\,+\,\Delta_{LB}H\right)\,\bm{n}
    \label{eq:force_bending}
\end{equation}
where $\Delta_{LB}$ is the Laplace-Beltrami operator. Details about Gauss curvature and local mean curvature are given in
the subsection \ref{section:Geometry}.
\subsection{Dimensionless quantities and relevant numbers}

Hereinafter, unless otherwise stated, the problem is now nondimensionalized by $x/R$, $\dot\epsilon t$ and $p/\eta\dot\epsilon$ respectively. The first dimensionless number which appears is the capillary number evaluating the viscous stress $\eta\dot\epsilon$ compared to the elastic one $G_s/R$:
\begin{equation}
    Ca\,=\,\frac{\eta \dot\epsilon R}{G_s}
\end{equation}
where $G_s$ is the one corresponding from the constitutive law that is used: $G^{SK}_s$, $G^{H}_s$ and $G^{NH}_s$. The larger $Ca$ is, the more deformed the capsule is. Thus $Ca$ is the small parameter driving the power expansion of all physical quantities. The second dimensionless number is the viscosity contrast which can be interpreted as the internal and external viscous stress ratio :
\begin{equation}
    \lambda = \frac{\eta^*}{\eta}
\end{equation}
The third one, the elastocapillary number $\Sigma$ is essential to understand some systems which couple the effect of surface tension and elasticity, the case of the wetting of a soft solid by a droplet is a prototype. $\Sigma$ is:
\begin{equation}
    \Sigma=\frac{\sigma}{G_s}
\end{equation}
The fourth one corresponds to the ratio of bending stress to tangent elastic ones:
\begin{equation}
    B=\frac{\kappa}{G_s\,R^2}
\end{equation}

From now on, the quantities used in this paper will be considered as dimensionally simplified. The non dimensional equations are then:
\begin{equation}
    r = 1+F\quad\quad \bm{n}= \frac{\bm{\nabla} r}{|\bm{\nabla} r|} = \frac{\bm{x}}{r} -r\bm{\nabla} F
    \label{eq:position}
\end{equation}
\begin{equation}
    \bm{v}^\infty = \left( \bm{E}+\bm{\Omega}\right)\cdot\bm{x}\text{ for }|\bm{x}|\rightarrow+\infty
    \label{eq:flow}
\end{equation}
\begin{equation}
    \left(\bm{v}+\bm{v}^\infty\right)\cdot\bm{n} = 0\label{eq:evolution-equation}
\end{equation}
\begin{equation}
    \Delta \bm{v} = \bm{\nabla} p\quad\quad\bm{\nabla}\cdot\bm{v} = 0
    \label{eq:StokesExt}
\end{equation}
\begin{equation}
    \Delta \bm{v}^* = \lambda \bm{\nabla} p^*\quad\quad\bm{\nabla}\cdot\bm{v}^* = 0
    \label{eq:StokesInt}
\end{equation}
\begin{equation}
    \left(\bm{\Pi}+\bm{\Pi}^\infty-\lambda\bm{\Pi}^*\right)\cdot\bm{n} = \frac{\bm{f}}{Ca}\label{eq:stress-continuity}
\end{equation}
\begin{equation}
    \bm{f}\,=\,\bm{f_{el}}\,+\,\bm{f_{\sigma}}\,+\,\bm{f_{\kappa}}
\end{equation}
where $\bm{f}$ is given by the equations (\ref{eq:fmembrane}), (\ref{eq:force_tension}) and (\ref{eq:force_bending}) divided by $G/R$ to ensure it is dimensionless. The forces expressions are written in appendix \ref{appendix:forces}.

\subsection{Numerical procedure}

The numerical computation of the fluid-structure interactions in the regime of Stokes flow allows to use the powerful method of boundary integral (\cite{pozrikidis2002practical,pozrikidis1992boundary}) and the linearity of Stokes equations. These integrals need special numerical treatment to deal with singularities. The method to treat such integrals consists in subtracting the singularities by using analytical integral identities, a well-documented method (\cite{pozrikidis1992boundary,pozrikidis2002practical}). We used a home-made numerical code based on Finite Element Method and Boundary Element Method (\cite{boedec2017isogeometric}) derived from a previous code (\cite{boedec20113d}). The calculation of the bending force has been numerically checked by comparison with a numerical code of another team  (\cite{guckenberger2016bending}) and by own computations of thin tubes where bending rigidity plays an essential role (\cite{boedec2013sedimentation}). Here, numerical computations are used to support the validation of our analytical results in their regimes of validity: $\lambda\,Ca\,<<\,1$. Indeed, we will use numerical simulations to confirm our theoretical results as it was done for the Chaffey and Brenner equation in the past (\cite{gounley2016influence}).

\section{Analytic procedure}

\subsection{Expansion to solid harmonics}

It is known that the small quantity defined inside the radius can be expanded to multiple orders. The first order gives information about the capsule deformation whilst the second order describes the orientation of the capsule according to a privileged axis chosen. In the frame of this paper, second order is adopted, meaning
\begin{equation}
    r = 1+F^{(1)}+F^{(2)}
\end{equation}
with $|F^{(2)}| \ll |F^{(1)}| \ll 1$ where the superscript denotes the order. Consequently, any quantity expressed at the interface should be expanded using Taylor expansion around the radius of the capsule. For example the velocity is
\begin{equation}
    \bm{v} = \bm{v}^{(1)}+\bm{v}^{(2)}+o(n)\text{ at r being the capsule radius}
\end{equation}
\begin{equation}
    \bm{v} = \bm{v}^{(1)}+\bm{v}^{(2)}+F^{(1)}\bm{x}\cdot\frac{\partial \bm{v}^{(1)}}{\partial \bm{x}}+o(n)\text{ at }r=1
\end{equation}
$o(n)$ is the higher orders.

By solving equations (\ref{eq:StokesExt}) and (\ref{eq:StokesInt}), \cite{lamb1924hydrodynamics} proposed the following expression for the velocity fields
\begin{equation}
    \bm{v} = \sum_{n=-\infty}^\infty \left[\bm{\nabla} \times \left(\bm{x} \chi^n\right)+\bm{\nabla} \phi^n + \frac{n+3}{2\left(n+1\right)\left(2n+3\right)}\bm{x}^2\bm{\nabla} p^n - \frac{n}{\left(n+1\right)\left(2n+3\right)}\bm{x}p^n\right]
\end{equation}
\begin{equation}
    v^* = \sum_{n=-\infty}^\infty \left[\bm{\nabla} \times \left(\bm{x} \chi^{n,*}\right)+\bm{\nabla} \phi^{n,*} + \frac{n+3}{2\left(n+1\right)\left(2n+3\right)}\bm{x}^2\bm{\nabla} p^{n,*} - \frac{n}{\left(n+1\right)\left(2n+3\right)}\bm{x}p^{n,*}\right]
\end{equation}
where $\chi$, $\chi^*$ are the rotational parts of the fields, $\phi$ and $\phi^*$ the anti-rotating parts and $p$, $p^*$ for the respective pressures. Those quantities are expanded on the basis of solid harmonics of order $n$ as described by \cite{frankel1968motion}. Since in a linear flow the harmonics are of order 2 and 4 for first and second orders, one can define the following quantities using tensorial notation
\begin{equation}
    p^{(1)} = T_{lm}^{(1)}\frac{\partial r^{-1}}{\partial_l \partial_m},\qquad \phi^{(1)} = S_{lm}^{(1)}\frac{\partial r^{-1}}{\partial_l \partial_m},\qquad \chi^{(1)} = C_{l}^{(1)}\frac{\partial r^{-1}}{\partial_l}
\end{equation}
and for harmonics of order 4
\begin{equation}
    p^{(2)} = T_{lm}^{(2)}\frac{\partial r^{-1}}{\partial_l \partial_m}+T_{lmpq}^{(2)}\frac{\partial r^{-1}}{\partial_l \partial_m \partial_p \partial_q},
\end{equation}
\begin{equation}
\phi^{(2)} = S_{lm}^{(2)}\frac{\partial r^{-1}}{\partial_l \partial_m}+S_{lmpq}^{(2)}\frac{\partial r^{-1}}{\partial_l \partial_m \partial_p \partial_q}
\end{equation}

The inner fluid follows the same decomposition
\begin{equation}
    p^{(1), *} = T_{lm}^{(1), *}r^5\frac{\partial r^{-1}}{\partial_l \partial_m},\qquad \phi^{(1), *} = S_{lm}^{(1), *}r^5\frac{\partial r^{-1}}{\partial_l \partial_m},\qquad \chi^{(1), *} = C_{l,}^{(1), *}r^3\frac{\partial r^{-1}}{\partial_l}
\end{equation}
\begin{equation}
    p^{(2), *} = T_{lm}^{(2), *}r^5\frac{\partial r^{-1}}{\partial_l \partial_m}+T_{lmpq}^{(2), *}r^9\frac{\partial r^{-1}}{\partial_l \partial_m \partial_p \partial_q}
\end{equation}
\begin{equation}
    \phi^{(2), *} = S_{lm}^{(2), *}r^5\frac{\partial r^{-1}}{\partial_l \partial_m}+S_{lmpq}^{(2), *}r^9\frac{\partial r^{-1}}{\partial_l \partial_m \partial_p \partial_q} 
\end{equation}\\

Based on those harmonics, one can see that the deformation tensor and the tensor from the reference state can also be expressed the same way on harmonics of order 2 and 4. Due to Laplace equation preserving the coherence of the harmonics, one need to impose the condition that the tensors used here are symmetrical and traceless. Taking \cite{barthes1981time} definition for the vector position and expanding to second order gives
\begin{equation}
\begin{split}
    \bm{x} &= \bm{X} + 3 \bm{K}^{(1)}\cdot\bm{X} + 3\left(\bm{F}^{(1)}-\bm{K}^{(1)}\right)\cdot\bm{X}\bm{X}\bm{X}/r^2 \\
    &+ 3 \bm{K}^{(2)}\cdot\bm{X} + 3\left(\bm{F}^{(2)}-\bm{K}^{(2)}\right)\cdot\bm{X}\bm{X}\bm{X}/r^2 + 105 \bm{K}^{(2)}\cdot\bm{X}\bm{X}\bm{X}/r^2\\
    &+ 105\left(\bm{F}^{(2)}-\bm{K}^{(2)}\right)\cdot\bm{X}\bm{X}\bm{X}\bm{X}\bm{X}/r^4-\frac{9}{2}\bm{K}^{(1)}\cdot\bm{K}^{(1)}\cdot \bm{X}\bm{X}\bm{X}/r^2\\
    &+\frac{9}{2}\left(\bm{K}^{(1)}\cdot\bm{X}\bm{X}\right)\left(\bm{K}^{(1)}\cdot\bm{X}\bm{X}\right)\bm{X}/r^4-\frac{6}{5}F^{(1)}_{lm}F^{(1)}_{lm}X_i
\end{split}
\end{equation}
Where the last term is added for the sake of volume conservation. This definition aims to obtain the same expression for the radius as in the problem of a clean droplet. Then at the interface, the radius is
\begin{equation}
    r = \sqrt{\bm{x}\cdot\bm{x}} = R\left(1+3 F^{(1)}_{lm}X_l X_m/r^2+3 F^{(2)}_{lm}X_l X_m/r^2+105 F^{(2)}_{lmpq}X_l X_mX_pX_q/r^4- \frac{6}{5}F^{(1)}_{lm}F^{(1)}_{lm}+o\left(n\right)\right)
\end{equation}

\subsection{Geometric quantities}
\label{section:Geometry}
The following section will be based on the formalism and definitions of \cite{frankel2004geometry}. From the radius expression, one can define the position vector at the interface
\begin{equation}
    \bm{r}_s = R\left(1+F\right)\frac{\bm{x}}{|\bm{x}|}
\end{equation}
where one can clearly see that $F$ is dimensionless and depends on the sphere angles. Consequently, one also defines
\begin{equation}
    F = F\left(\frac{\bm{X}}{r}\right) = F\left(\bm{\xi}\right)
\end{equation}
$\bm{\xi}$ is the set of curvilinear coordinates. To prevent any confusions, curvilinear coordinates will come with greek indices such that
\begin{equation}
    \xi^\mu = \left(\xi^1, \xi^2\right)
\end{equation}
This allows us to determine the tangent vectors
\begin{equation}
    \bm{t}_\mu = \partial_\mu \bm{r}_s
\end{equation}
the induced metric is then obtained
\begin{equation}
    h_{\mu\nu} = \bm{t}_\mu\cdot \bm{t}_\nu
\end{equation}
also the normal vector is obtained by the postulate that we can build a basis involving the normal as $\left(\bm{t}_1, \bm{t}_2, \bm{n}\right)$ which leads to the formula
\begin{equation}
    \bm{n} = \frac{\bm{t}_1\times \bm{t}_2}{|\bm{t}_1\times \bm{t}_2|}
\end{equation}
and has the local value
\begin{equation}
\begin{split}
    n_i &= \frac{X_i}{r}+\frac{-6F^{(1)}_{il} + 3K^{(1)}_{il}}{r}X_l+\frac{6F^{(1)}_{lm} - 3K^{(1)}_{lm}}{r^3}X_lX_mX_i\\
    &+\frac{-6F^{(2)}_{il} + 3K^{(2)}_{il}}{r}X_l+\frac{6F^{(2)}_{lm} - 3K^{(2)}_{lm}}{r^3}X_lX_mX_i\\
    &+\frac{-420 F^{(2)}_{impq} + 105 K^{(2)}_{impq}}{r^3}X_aX_bX_c+\frac{420 F^{(2)}_{lmpq} - 105 K^{(2)}_{lmpq}}{r^5}X_lX_mX_pX_qX_i\\
    &+\frac{18 F^{(1)}_{la}K^{(1)}_{ai}}{r}X_l + \frac{18 F^{(1)}_{mp}F^{(1)}_{il} - 18F^{(1)}_{il}K^{(1)}_{mp}-27F^{(1)}_{lm}K^{(1)}_{ip}}{r^3}X_lX_mX_p\\
    &+\frac{-18 F^{(1)}_{la}F^{(1)}_{am}-\frac{9}{2}K^{(1)}_{la}K^{(1)}_{am}}{r^3}X_lX_mX_i+\frac{27 F^{(1)}_{lm}K^{(1)}_{pq}+\frac{9}{2}K^{(1)}_{lm}K^{(1)}_{pq}}{r^5}X_lX_mX_pX_qX_i
\end{split}
\end{equation}
Finally, the second fundamental form is
\begin{equation}
    b_{\mu\nu} = -\bm{n}\cdot \partial_\nu \bm{t}_\mu
\end{equation}
This quantity is the one encoding the curvature of the object. The Gauss curvature $K_g$ and the local mean curvature $H$ are obtained
\begin{equation}
    K_g = det\left(b^\mu_\nu\right),\qquad H = \frac{1}{2}h^{\mu\nu}b_{\nu\mu}
\end{equation}
Their local values are
\begin{equation}
\begin{split}
    H &= \frac{1}{r}+\frac{\frac{6}{5}F^{(1)}_{lm}K^{(1)}_{lm}+18 F^{(1)}_{lm}K^{(1)}_{lm}}{r}+\frac{6 F^{(1)}_{lm}+6 F^{(2)}_{lm}-72F^{(1)}_{la}K^{(1)}_{am}}{r^3}X_lX_m\\
    &+\frac{-45 F^{(1)}_{lm}F^{(1)}_{pq}+126F^{(1)}_{lm}K^{(1)}_{pq}+945F^{(2)}_{lmpq}}{r^5}X_lX_mX_pX_q
\end{split}
\end{equation}

\begin{equation}
\begin{split}
    K_g &= \frac{1}{r^2}+\frac{-18F^{(1)}_{lm}F^{(1)}_{lm}+36F^{(1)}_{lm}K^{(1)}_{lm}}{r^2}\\
    &+\frac{12 F^{(1)}_{lm}+12F^{(2)}_{lm}+36F^{(1)}_{la}F^{(1)}_{am}-144F^{(1)}_{la}K^{(1)}_{am}}{r^4}X_lX_m\\
    &+\frac{-63 F^{(1)}_{lm}F^{(1)}_{pq}+252F^{(1)}_{lm}K^{(1)}_{pq}+1890F^{(2)}_{lmpq}}{r^6}X_lX_mX_pX_q
\end{split}
\end{equation}
One recovers the local mean curvature from \cite{frankel1968motion} for a clean droplet by setting the tensor $\bm{K}=0$ and by multiplying $H$ with a factor $2$. This last difference comes from the used convention for $H$.
\subsection{Resolution at first order}

The study will deal with steady state, meaning $\partial x/\partial t=0$. To solve equations (\ref{eq:velocity-continuity}), (\ref{eq:evolution-equation}) and (\ref{eq:stress-continuity}), one multiplies them by $\bm{x}$, $\bm{x}\bm{x}\bm{x}$ or even more and integrate over an angular sphere volume $d\Omega$
\begin{equation}
    \int x_ix_jd\Omega = \frac{4\pi}{3}\delta_{ij}
\end{equation}
\begin{equation}
    \int x_ix_jx_lx_md\Omega = \frac{4\pi}{15}\left(\delta_{ij}\delta_{lm}+\delta_{il}\delta_{jm}+\delta_{im}\delta_{jl}\right)
\end{equation}
and so on for higher orders. One also has to take into account the application of the symmetrization operators
\begin{equation}
    Sd_2\left[T_{lm}\right] = \frac{1}{2}\left(T_{lm}+T_{ml}-\frac{2}{3}\delta_{lm}T_{jj}\right)
\end{equation}
\begin{equation}
\begin{split}
    Sd_4\left[T_{ijkl}\right] &= \frac{1}{6}\left(T_{ijkl}+\text{permutations}\right)-\frac{2}{21}Sd_2\left[T_{pmmq}\right]\left(\delta_{ij}\delta_{kp}\delta_{ql}+\text{permutations}\right)\\
    &-\frac{2}{45}T_{mpmp}\left(\delta_{ij}\delta_{kl}+\text{permutations}\right)
    \end{split}
\end{equation}
as well as the commutator
\begin{equation}
    \left[\bm{E}, \bm{F}\right] = \bm{E}\cdot\bm{F} - \bm{F}\cdot\bm{E}
\end{equation}
At first order, using this method, the integration of equation (\ref{eq:evolution-equation}) gives the following expressions
\begin{equation}
    \frac{3}{15}T_{ij}^{(1)}+\frac{1}{3}E_{ij} = 0
\end{equation}
\begin{equation}
    -9S_{ij}^{(1)}+\frac{3}{2}T_{ij}^{(1)}+E_{ij} = 0
\end{equation}
Where the quantities from the velocity field are determined earlier and depends on the constitutive law. Solving both equations can be tedious at first glance. By applying the method from \cite{barthes1985role} the following equation can be written
\begin{equation}
    0 = M X + Q\label{eq:matrix}
\end{equation}
where $M$ denotes the matrix containing the quantities dependent of $F$ and $K$ while $Q$ contains the boundary conditions. For example, using Skalak law without bending and surface tension forces, one would have
\begin{equation}
    \begin{pmatrix}
    0\\
    0
    \end{pmatrix} = \begin{pmatrix}
        \frac{-4\left(1+2C\right)\left(4+\lambda\right)}{Ca\left(3+2\lambda\right)\left(16+19\lambda\right)} && \frac{6\left(-5\lambda+2C\left(4+\lambda\right)\right)}{Ca\left(3+2\lambda\right)\left(16+19\lambda\right)}\\
        \frac{4\left(1+2C\right)\left(8+7\lambda\right)}{Ca\left(3+2\lambda\right)\left(16+19\lambda\right)} && \frac{-2\left(48+47C+6C\left(8+7\lambda\right)\right)}{Ca\left(3+2\lambda\right)\left(16+19\lambda\right)}
    \end{pmatrix}\begin{pmatrix}
    \bm{F}^{(1)}\\
    \bm{K}^{(1)}
    \end{pmatrix}+\begin{pmatrix}
        \frac{5}{3\left(3+2\lambda\right)}\bm{E}\\
        \frac{5}{3\left(3+2\lambda\right)}\bm{E}
    \end{pmatrix}
\end{equation}
Using the well-known diagonalization formula $M = P D P^{-1}$ one can easily solve this system of equations.\\

\subsection{Second order}

The second order consists in solving the following equations
\begin{equation}
    \bm{v}^{(2)}+F^{(1)}\bm{x}\cdot\bm{\nabla} \bm{v}^{(1)} = \bm{v}^{(2),*}+F^{(1)}\bm{x}\cdot\bm{\nabla} \bm{v}^{(2),*}\label{eq:second_velocity_continuity}
\end{equation}
\begin{equation}
    \left(\bm{\Pi}^{(2)}-\lambda\bm{\Pi}^{(2),*}\right)\cdot\bm{n}^{(0)}+F^{(1)}\bm{x}\cdot\bm{\nabla}\left(\bm{\Pi}^{(1)}-\lambda\bm{\Pi}^{(1),*}\right)\cdot\bm{n}^{(0)}+\left(\bm{\Pi}^{(1)}-\lambda\bm{\Pi}^{(1),*}\right)\cdot\bm{n}^{(1)} = \bm{f}^{(2)}\label{eq:second_stress_continuity}
\end{equation}
\begin{equation}
    \bm{v}^{(2)}\cdot\bm{n}^{(0)}+F^{(1)}\bm{x}\cdot\bm{\nabla} \bm{v}^{(1)}\cdot\bm{n}^{(0)}+\bm{v}^{(1)}\cdot\bm{n}^{(1)} = 0\label{eq:second_evolution_equation}
\end{equation}
The only difference is that one has to proceed by solving the equations for each harmonics; so order 2 and 4. Using the following local decomposition
\begin{equation}
    E_{il}F_{lj} = Sd_2\left[E_{il}F_{lj}\right] + \left[ E_{il}F_{lj} - F_{il}E_{lj} \right]
\end{equation}
for each corresponding tensors, the evolution equation gives the expressions for harmonics of order 2
\begin{equation}
\begin{split}
    &\frac{1}{5}T_{ij}^{(2)}-\frac{12}{7}Sd_2\left[T_{il}^{(1)}F_{lj}^{(1)}\right]+\frac{9}{35}Sd_2\left[T_{il}^{(1)}K_{lj}^{(1)}\right]+\frac{432}{35}Sd_2\left[S_{il}^{(1)}F_{lj}^{(1)}\right]\\
    & +\frac{36}{35}Sd_2\left[S_{il}^{(1)}K_{lj}^{(1)}\right]-\frac{4}{5}Sd_2\left[E_{il}F_{lj}^{(1)}\right]+\frac{3}{5}Sd_2\left[E_{il}K_{lj}^{(1)}\right]-\left(\Omega_{il}F_{lj}^{(1)}-F_{il}\Omega_{lj}\right) = 0
\end{split}
\end{equation}
\begin{equation}
\begin{split}
    &-\frac{6}{5}S_{ij}^{(2)}+\frac{1}{5}T_{ij}^{(2)}-\frac{24}{35}Sd_2\left[T_{il}^{(1)}F_{lj}^{(1)}\right]+\frac{216}{35}Sd_2\left[S_{il}^{(1)}F_{lj}^{(1)}\right]+\frac{36}{35}Sd_2\left[S_{il}^{(1)}K_{lj}^{(1)}\right]\\
    &-\frac{4}{35}Sd_2\left[E_{il}F_{lj}^{(1)}\right]+\frac{6}{35}Sd_2\left[E_{il}K_{lj}^{(1)}\right]-\frac{2}{5}\left(\Omega_{il}F_{lj}^{(1)}-F_{il}\Omega_{lj}\right) = 0
\end{split}
\end{equation}
The expressions for harmonics of order 4 are
\begin{equation}
\begin{split}
    &\frac{10}{21}T_{ijkl}^{(2)}+\frac{2}{105}\left(432Sd_4\left[S_{ij}^{(1)}F_{kl}^{(1)}\right]-54Sd_4\left[S_{ij}^{(1)}K_{kl}^{(1)}\right]\right)-\frac{18}{35}Sd_4\left[T_{ij}^{(1)}F_{kl}^{(1)}\right]\\
    &+\frac{2}{105}\left(27Sd_4\left[E_{ij}F_{kl}^{(1)}\right]-9Sd_4\left[E_{ij}K_{kl}^{(1)}\right]\right)= 0
\end{split}
\end{equation}
\begin{equation}
\begin{split}
    &-\frac{40}{3}S_{ijkl}^{(2)}+\frac{20}{21}T_{ijkl}^{(2)}+\frac{4}{105}\left(96Sd_4\left[S_{ij}^{(1)}F_{kl}^{(1)}\right]-12Sd_4\left[S_{ij}^{(1)}K_{kl}^{(1)}\right]\right)\\
    & -\frac{8}{35}Sd_4\left[T_{ij}^{(1)}F_{kl}^{(1)}\right] + \frac{4}{105}\left(6Sd_4\left[E_{ij}F_{kl}^{(1)}\right]-2Sd_4\left[E_{ij}K_{kl}^{(1)}\right]\right)= 0
\end{split}
\end{equation}
The solutions are straightforward once again by using matrix diagonalization. The solutions found are of order 2. They allow to quantify the orientation of the capsule and will be discussed in the next chapter. The values for the matrices are given in appendix \ref{appendix:matrix}.

\section{Results and discussion}

In a  linear flow, a capsule deforms into an ellipsoid characterized by three semi-axis lengths called here $(L,S,W)$ where $L$ is the longest one at the leading order. \cite{taylor1934formation} defined the deformation of a spherical droplet in linear flows. The so-called Taylor parameter $D$ reads
\begin{equation}
    D = \frac{L-S}{L+S}
\end{equation}
To support any experimental analysis as in \cite{de2015stretching}, the expressions of $(L,S,W)$ in shear and planar extensional flows are provided in the following. These expressions lead to the definition of three Taylor parameters: $D_{12}$, $D_{13}$, $D_{23}$. In planar extensional flow, these quantities are the cross-sections in each direction. In shear flow, extra care is required as the cross-section deformations are different from the true deformation considered in this paper as discussed by \cite{barthes1985role}. In this case to characterize the equivalent ellipsoid, first, the inclination angle $\phi_{TT}$ is determined by the orientation of the longest semi-axis $L$: maximum of the equation (\ref{eq:position}). The two others semi-axis are provided by rotations in spherical coordinates:\\
\begin{equation}
    L = r\left(\theta = \frac{\pi}{2},\ \phi = \phi_{TT}\right)
\end{equation}
\begin{equation}
    S = r\left(\theta = \frac{\pi}{2},\ \phi = \phi_{TT}+\frac{\pi}{2}\right)
\end{equation}
\begin{equation}
    W = r\left(\theta = 0\right)
\end{equation}
Using taylor expansions for trigonometric functions the orientation angle should look like
\begin{equation}
    \phi_{TT} = \frac{\pi}{4} - \delta\left(Ca\right)+o\left(Ca^2\right)
\end{equation}
where $\delta$ is a function of the capillary number and the viscosity contrast coming from the second order theory.\\

\subsection{Deformation and orientation under shear flow}
In shear flow, the velocity given by (\ref{eq:flow}) is defined by the following quantities
\begin{equation}
    \bm{E} = \begin{pmatrix}
        0 && 1/2 && 0\\
        1/2 && 0 && 0\\
        0 && 0 && 0
    \end{pmatrix},\qquad \bm{\Omega} = \begin{pmatrix}
        0 && 1/2 && 0\\
        -1/2 && 0 && 0\\
        0 && 0 && 0
    \end{pmatrix}
\end{equation}
\subsubsection{The Skalak model}

Solving equations (\ref{eq:velocity-continuity}), (\ref{eq:evolution-equation}) and (\ref{eq:stress-continuity}) one obtains
\begin{equation}
    F_{ij}^{(1)} = E_{ij}\frac{5 (3 C+2)}{3 \left(C \left(36B
+6 \Sigma +4\right)+30B+5 \Sigma +2\right)}Ca
\end{equation}
\begin{equation}
    K_{ij}^{(1)} = E_{ij}\frac{5 \left(6B+\Sigma +4 C+2\right)}{6 \left(C \left(36B+6 \Sigma +4\right)+30B+5
   \Sigma +2\right)}Ca
\end{equation}
For $C=1$ and a factor $3$ depending on the definition of $G^{SK}$ one retrieves the results from \cite{barthes1981time} in the stationary state. The tensors $F$ and $K$ are proportional to the capillary number which satisfies the requirement that those tensors are small according to the definition of the radius.
The matrices $M$ and $Q$ are determined by solving equations (\ref{eq:second_velocity_continuity}), (\ref{eq:second_stress_continuity}) and (\ref{eq:second_evolution_equation}). Their values are given in appendix \ref{appendix:matrix}.\\

\begin{figure}
    \centering
    \includegraphics[width=1\linewidth]{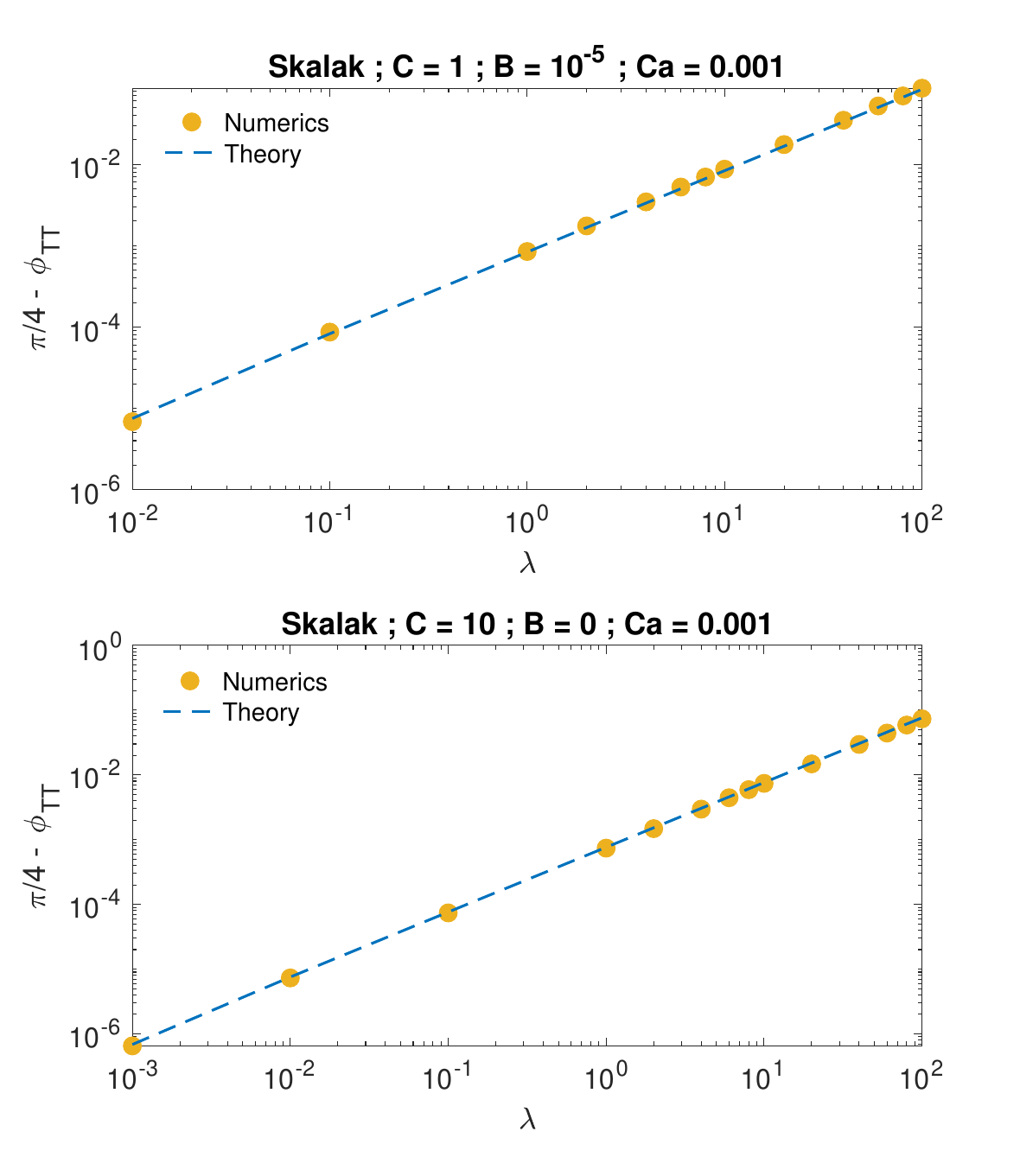}
    \caption{Comparison between the numerical results and the theoretical equation (\ref{eq:phiSk}) in the case of a capsule obeying to the Skalak constitutive law and with a viscosity contrast $\lambda$ : deviation from $\pi/4$ of the angle $\phi_{TT}$ of the capsule's orientation with the direction of the  shear flow. The results are valid in the limit $\lambda\,Ca\,<<\,1$.}
    \label{fig:angleSk}
\end{figure}
In shear flow, the angle $\phi_{TT}$ is
\begin{equation}
    \phi^{Sk} = \frac{\pi}{4}-\frac{(2 \lambda +3) (3 C+2)}{2 (4 C+2)}\frac{4 C+2}{C \left(36B+6 \Sigma +4\right)+30B+5 \Sigma +2}Ca\,+\,o(Ca^2)
    \label{eq:phiSk}
\end{equation}
Which gives the following semi-axis lengths
\begin{equation}
    L = 1+\frac{5\left(2+3C\right)}{4\left(1+2C\right)}\frac{4 C+2}{C \left(36B+6 \Sigma +4\right)+30B+5 \Sigma +2}Ca+o\left(Ca^3\right)
\end{equation}
\begin{equation}
    S = 1-\frac{5\left(2+3C\right)}{4\left(1+2C\right)}\frac{4 C+2}{C \left(36B+6 \Sigma +4\right)+30B+5 \Sigma +2}Ca+o\left(Ca^3\right)
\end{equation}
\begin{equation}
    W = 1+o\left(Ca^3\right)
\end{equation}
and the deformations in the 12, 13, and 23 planes
\begin{equation}
    D^{shear}_{12} = \frac{5\left(2+3C\right)}{4\left(1+2C\right)}\frac{4 C+2}{C \left(36B+6 \Sigma +4\right)+30B+5 \Sigma +2}Ca+o\left(Ca^3\right)
\end{equation}
\begin{equation}
    D^{shear}_{13} = -D^{shear}_{23} = \frac{1}{2}D^{shear}_{12}
\end{equation}
A comparison between the numerical simulations and our theoretical results is shown in figure \ref{fig:angleSk}. Only the variation of the orientation angle (\ref{eq:phiSk}) is plotted to highlight the excellent agreement.\\

\subsubsection{The Hooke model}
The Hooke model is quite similar and leads to the following tensors
\begin{equation}
    F_{ij}^{(1)} = E_{ij}\frac{5 (\nu+2)}{3 \left(\nu \left(6B+\Sigma +2\right)+30B+5 \Sigma +2\right)}Ca
\end{equation}
\begin{equation}
    K_{ij}^{(1)} = E_{ij}\frac{5 \left(-\nu \left(6B+\Sigma -2\right)+6B+\Sigma +2\right)}{6 \left(\nu \left(6B+\Sigma
   +2\right)+30B+5 \Sigma +2\right)}Ca
\end{equation}

In shear flow the angle and the deformations are
\begin{equation}
    \phi^{H} = \frac{\pi}{4}-\frac{ (2 \lambda +3) (\nu+2)}{4 (\nu+1)}\frac{2 (\nu+1)}{\nu(6B+\Sigma+2)+30B+5\Sigma+2} Ca\,+\,o(Ca^2)
\end{equation}
\begin{equation}
    L = 1+\frac{5 (\nu+2)}{4 ( \nu+1)}\frac{2 (\nu+1)}{\nu(6B+\Sigma+2)+30B+5\Sigma+2}Ca+o\left(Ca^3\right)
\end{equation}
\begin{equation}
    S = 1-\frac{5 (\nu+2)}{4 ( \nu+1)}\frac{2 (\nu+1)}{\nu(6B+\Sigma+2)+30B+5\Sigma+2}Ca+o\left(Ca^3\right)
\end{equation}
\begin{equation}
    W = 1+o\left(Ca^3\right)
\end{equation}
\begin{equation}
    D^{shear}_{12} = \frac{5 (\nu+2)}{4 ( \nu+1)}\frac{2 (\nu+1)}{\nu(6B+\Sigma+2)+30B+5\Sigma+2}Ca+o\left(Ca^3\right)
\end{equation}
\begin{equation}
    D^{shear}_{13} = -D^{shear}_{23} = \frac{1}{2}D^{shear}_{12}
\end{equation}
The expressions of the deformation in the cases of the Hooke's model and the Skalak's model are equivalent in the case $\nu = C/\left(1+C\right)$.\\

\begin{figure}
    \centering
    \includegraphics[width=1\linewidth]{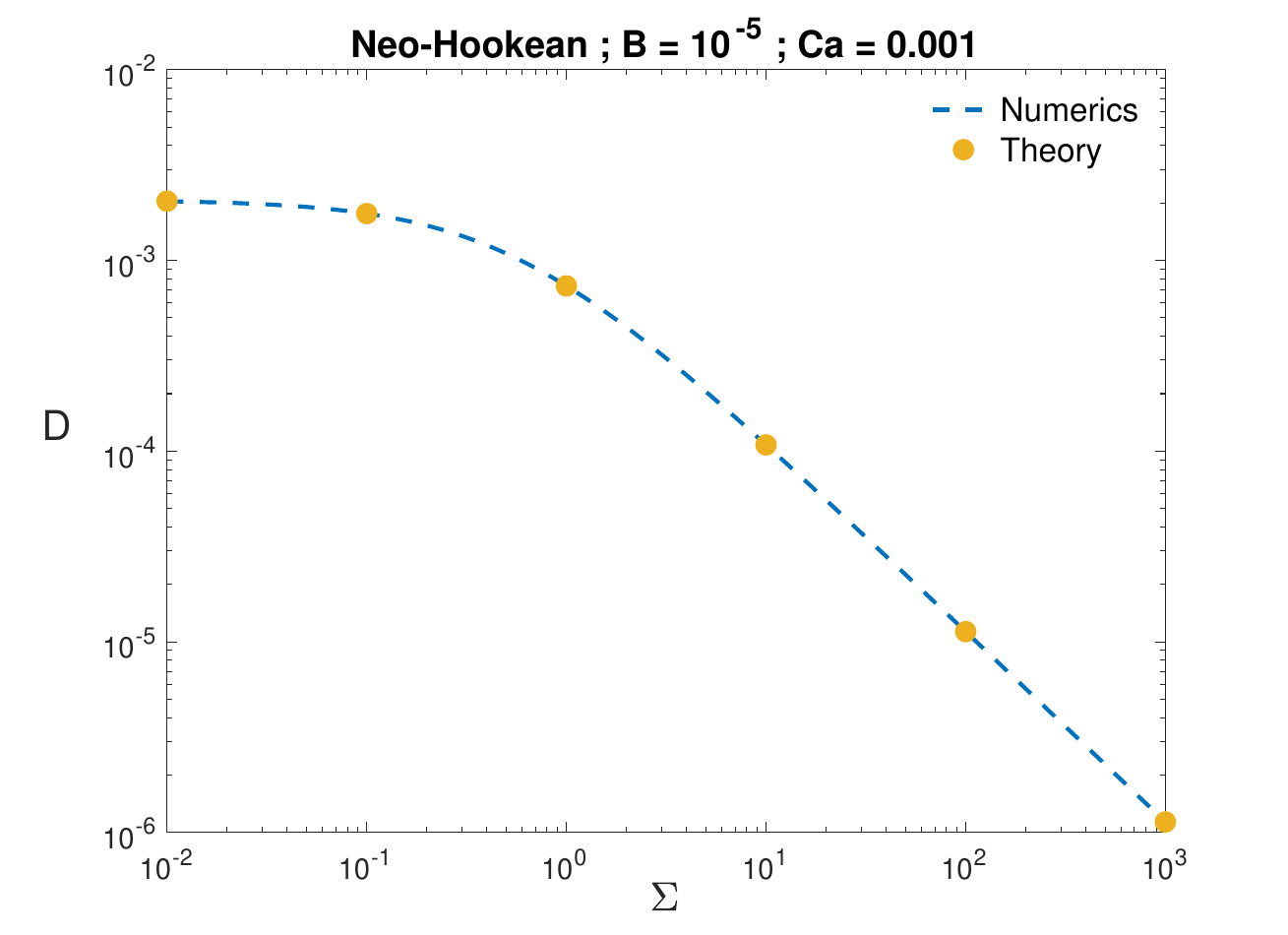}
    \caption{Comparison between the numerical results and the theoretical equation (\ref{eq:DNH}) of the deformation of a Neo-Hookean capsule with the surface elastocapillary number $\Sigma$ under shear flow. The results are determined in the limit $\lambda\,Ca\,<<\,1$. The range of $\Sigma$ is chosen to valid fully the agreement.}
    \label{fig:DSigma}
\end{figure}

\subsubsection{The Neo-Hookean}
Here, the tensors are
\begin{equation}
    F_{ij}^{(1)} = E_{ij}\frac{25}{198B+33 \Sigma +18}Ca
\end{equation}
\begin{equation}
    K_{ij}^{(1)} = E_{ij}\frac{5 \left(6B+\Sigma +6\right)}{6 \left(66B+11 \Sigma +6\right)}Ca
\end{equation}

The Neo-Hookean case follows the same structure being
\begin{equation}
    \phi^{NH} = \frac{\pi}{4}-\frac{5(2 \lambda +3)}{12}\frac{6}{66B+11 \Sigma +6}Ca\,+\,o(Ca^2)
\end{equation}
\begin{equation}
    L = 1+\frac{25}{12}\frac{6}{66B+11 \Sigma +6}Ca+o\left(Ca^3\right)
\end{equation}
\begin{equation}
    S = 1-\frac{25}{12}\frac{6}{66B+11 \Sigma +6}Ca+o\left(Ca^3\right)
\end{equation}
\begin{equation}
    W = 1+o\left(Ca^3\right)
\end{equation}
\begin{equation}
    D^{shear}_{12} = \frac{25}{12}\frac{6}{66B+11 \Sigma +6}Ca+o\left(Ca^3\right)
    \label{eq:DNH}
\end{equation}
\begin{equation}
    D^{shear}_{13} = -D^{shear}_{23} = \frac{1}{2}D^{shear}_{12}
\end{equation}

\begin{figure}
    \centering
    \includegraphics[width=1\linewidth]{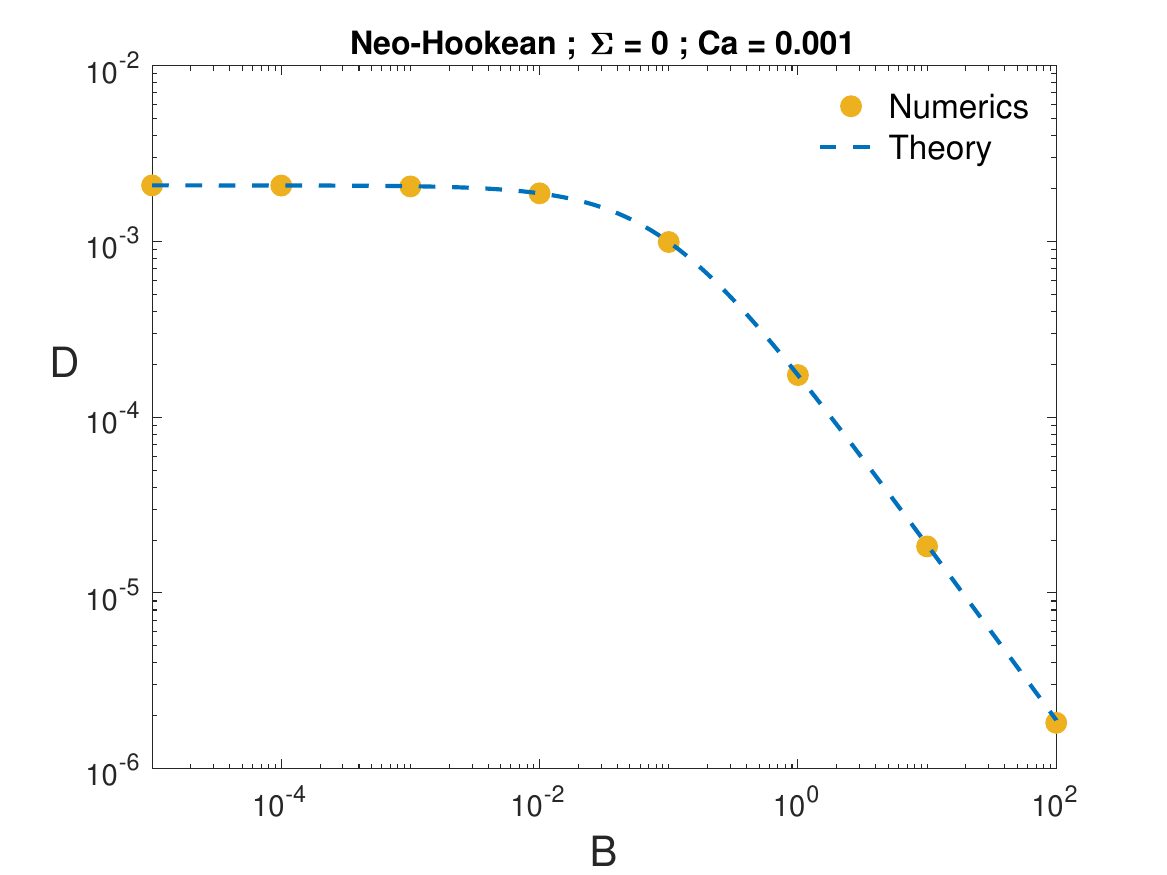}
    \caption{Comparison between the numerical results and the theoretical equation (\ref{eq:DNH}) of the deformation of a Neo-Hookean capsule with dimensionless bending rigidity $B$ under shear flow. The results are determined in the limit $\lambda\,Ca\,<<\,1$. The range of $B$ is chosen to valid fully the agreement.}
    \label{fig:DB}
\end{figure}

A comparison between the numerical simulations and our theoretical results is shown in figure \ref{fig:DSigma} versus the elastocapillary number $\Sigma$ and figure \ref{fig:DB} versus the dimensionless bending rigidity $B$. In both cases, the agreement is excellent even in the asymptotic limits of high $\Sigma$ or high $B$.

\subsection{Deformation under planar extensional flow}
In planar extensional flow, the velocity given by equation (\ref{eq:flow}) is defined by the following quantities
\begin{equation}
    \bm{E} = \begin{pmatrix}
        1 && 0 && 0\\
        0 && -1 && 0\\
        0 && 0 && 0
    \end{pmatrix},\qquad \bm{\Omega} = 0
\end{equation}
which lead to the same kind of results as under shear flow. The deformation differs by a factor 2 up to the order 2 while the longest axis $L$ of the capsule is aligned with the x-axis. The deformations $D^{ext}$ under planar extensional flow read:\\

$\bullet$ Skalak model:
\begin{equation}
    D^{ext}_{12} = \frac{5\left(2+3C\right)}{2\left(1+2C\right)}\frac{4 C+2}{C \left(36B+6 \Sigma +4\right)+30B+5 \Sigma +2}Ca+o\left(Ca^3\right)
\end{equation}\\

$\bullet$ Hooke model:
\begin{equation}
    D^{ext}_{12} = \frac{5 (\nu+2)}{2 ( \nu+1)}\frac{2 (\nu+1)}{\nu(6B+\Sigma+2)+30B+5\Sigma+2}Ca+o\left(Ca^3\right)
\end{equation}\\

$\bullet$ Neo-Hookean model:
\begin{equation}
    D^{ext}_{12} = \frac{25}{6}\frac{6}{66B+11 \Sigma +6}Ca+o\left(Ca^3\right)
    \label{eq:DNH}
\end{equation}

\section{Conclusion}
Throughout this paper, the stationary deformation and orientation of an initially spherical capsule has been studied theoretically in linear flows, namely under shear and planar extensional flows. All the quantities (shape, velocity, pressure, stress, membrane mechanical response) have been expanded up to the second order in the capillary number in the framework of three surface constitutive laws: Skalak, Hooke and Neo-Hookean models. The results are valid in the limit of small deformations, namely $ \lambda Ca << 1$. The surface tension and the bending modulus are taken into account as they play an essential role in the cases of low polymerization (porous membrane) and oil-in-water (or water-in-oil) capsule or when wrinkling or buckling appears. Indeed, to study stability or to make inverse analysis to gain insight in the membrane material's properties, the steady state has to be well defined. Without surface tension and bending rigidity, the previous results from \cite{barthes1981time} for a Skalak law and from \cite{lac2005deformation} for a Hooke law determined at the first order are recovered. We demonstrate that the contribution of the second order to deformation is zero whatever the constitutive law. It is also the case for a droplet as demonstrated by \cite{vlahovska2009small}. Thus, the non linear mechanical behavior will appear at the third order. This explains why the experimental measurements of the deformation under planar extensional flow exhibit a clear linear response over a large range of capillary number allowing an efficient inverse analysis (determination of the shear elastic modulus) without ambiguity. The expressions of the angle of inclination of a capsule under shear flow are also determined in the general case with surface tension and bending rigidity. They are analog of the Chaffey-Brenner expression for a droplet (\cite{chaffey1967second}). Contrary to the deformation, the orientation depends on the viscosity contrast as in the case of a droplet. In the limit of high surface elastocapillary number $\Sigma$ or dimensionless bending modulus $B$, the deformation and the orientation do not depend on the shear elastic modulus. These asymptotic limits are useful to establish the relevance of the theoretical results which indeed, are in excellent agreement with the numerical ones. Finally, this allows to ensure that these analytical expressions can also be useful to valid some new numerical developments.

\section{acknowledgments}
Paul Regazzi received
support from the french government under the France 2030 investment plan, as part of the
Initiative d’Excellence d’Aix-Marseille Université – AMIDEX and the CNRS through PhD grant. We also acknowledge for funding support from the
French Agence Nationale de la Recherche in the
framework of project SOCA, ANR-24-CE06-6245, the National Centre for Space Studies, CNES. Centre de Calcul Intensif d’Aix-Marseille is acknowledged for granting access to its high performance computing resources.
\clearpage
\bibliographystyle{apalike}
\bibliography{biblio_these}
\clearpage
\appendix
\section{Expression of local forces}\label{appendix:forces}
\subsection{Elastic force}
The elastic force based on the constitutive law takes the following form
\begin{equation}
    \begin{split}
        \frac{f_i^{SK}}{G_s^{SK}} &= \frac{480 C F^{(1)}_{bc} F^{(1)}_{i,a} X_b X_c X_a}{r^4}-\frac{156 F^{(1)}_{bc} F^{(1)}_{i,a} X_b X_c
   X_a}{r^4}+\frac{1680 C F^{(2)}_{i a b c} X_b X_c X_a}{r^4}+\frac{840 F^{(2)}_{i a b c} X_b X_c
   X_a}{r^4}\\
   &-\frac{900 C F^{(1)}_{i,a} K^{(1)}_{bc} X_b X_c X_a}{r^4}+\frac{126 F^{(1)}_{i,a} K^{(1)}_{bc}
   X_b X_c X_a}{r^4}+\frac{1188 C K^{(1)}_{bc} K^{(1)}_{i,a} X_b X_c X_a}{r^4}+\frac{126 K^{(1)}_{bc}
   K^{(1)}_{i,a} X_b X_c X_a}{r^4}\\
   &-\frac{648 C F^{(1)}_{a b} K^{(1)}_{i c} X_b X_c X_a}{r^4}-\frac{90
   F^{(1)}_{a b} K^{(1)}_{i c} X_b X_c X_a}{r^4}-\frac{4200 C K^{(2)}_{i a b c} X_b X_c X_a}{r^4}-\frac{3990
   K^{(2)}_{i a b c} X_b X_c X_a}{r^4}\\
   &-\frac{48 C F^{(1)}_{a b} X_b X_i X_a}{r^4}-\frac{24 F^{(1)}_{a b} X_b
   X_i X_a}{r^4}-\frac{48 C F^{(2)}_{a b} X_b X_i
   X_a}{r^4}-\frac{24 F^{(2)}_{a b} X_b X_i X_a}{r^4}+\frac{72 C K^{(1)}_{a b} X_b X_i X_a}{r^4}\\
   &+\frac{48
   K^{(1)}_{a b} X_b X_i X_a}{r^4}-\frac{72 C F^{(1)}_{a c} K^{(1)}_{bc} X_b X_i X_a}{r^4}-\frac{360
   F^{(1)}_{a c} K^{(1)}_{bc} X_b X_i X_a}{r^4}+\frac{108 C K^{(1)}_{a c} K^{(1)}_{bc} X_b X_i
   X_a}{r^4}\\
   &+\frac{288 K^{(1)}_{a c} K^{(1)}_{bc} X_b X_i X_a}{r^4}+\frac{72 C K^{(2)}_{a b} X_b X_i
   X_a}{r^4}+\frac{48 K^{(2)}_{a b} X_b X_i X_a}{r^4}-\frac{972 C F^{(1)}_{a b} F^{(1)}_{c d} X_b X_c X_d X_i
   X_a}{r^6}\\
   &+\frac{54 F^{(1)}_{a b} F^{(1)}_{c d} X_b X_c X_d X_i X_a}{r^6}-\frac{2520 C F^{(2)}_{a b c d} X_b
   X_c X_d X_i X_a}{r^6}-\frac{1260 F^{(2)}_{a b c d} X_b X_c X_d X_i X_a}{r^6}\\
   &+\frac{2808 C F^{(1)}_{a b}
   K^{(1)}_{c d} X_b X_c X_d X_i X_a}{r^6}+\frac{252 F^{(1)}_{a b} K^{(1)}_{c d} X_b X_c X_d X_i
   X_a}{r^6}-\frac{1944 C K^{(1)}_{a b} K^{(1)}_{c d} X_b X_c X_d X_i X_a}{r^6}\\
   &-\frac{216 K^{(1)}_{a b}
   K^{(1)}_{c d} X_b X_c X_d X_i X_a}{r^6}+\frac{6300 C K^{(2)}_{a b c d} X_b X_c X_d X_i
   X_a}{r^6}+\frac{5040 K^{(2)}_{a b c d} X_b X_c X_d X_i X_a}{r^6}+\frac{24 C F^{(1)}_{i,a}
   X_a}{r^2}\\
   &+\frac{12 F^{(1)}_{i,a} X_a}{r^2}+\frac{72 C F^{(1)}_{a b} F^{(1)}_{i b} X_a}{r^2}+\frac{36
   F^{(1)}_{a b} F^{(1)}_{i b} X_a}{r^2}+\frac{24 C F^{(2)}_{i,a} X_a}{r^2}+\frac{12 F^{(2)}_{i,a}
   X_a}{r^2}-\frac{36 C F^{(1)}_{i b} K^{(1)}_{a b} X_a}{r^2}\\
   &-\frac{18 F^{(1)}_{i b} K^{(1)}_{a b}
   X_a}{r^2}-\frac{36 C K^{(1)}_{i,a} X_a}{r^2}-\frac{30 K^{(1)}_{i,a} X_a}{r^2}-\frac{108 C F^{(1)}_{a b}
   K^{(1)}_{i b} X_a}{r^2}+\frac{126 F^{(1)}_{a b} K^{(1)}_{i b} X_a}{r^2}+\frac{54 C K^{(1)}_{a b} K^{(1)}_{i b} X_a}{r^2}\\
   &-\frac{171 K^{(1)}_{a b} K^{(1)}_{i b} X_a}{r^2}-\frac{36 C K^{(2)}_{i,a} X_a}{r^2}-\frac{30
   K^{(2)}_{i,a} X_a}{r^2}+\frac{48 C F^{(1)}_{a b} F^{(1)}_{a b} X_i}{5 r^2}-\frac{18 C K^{(1)}_{a b}
   K^{(1)}_{a b} X_i}{r^2}
    \end{split}
\end{equation}
\begin{equation}
    \begin{split}
        \frac{f_i^{H}}{G_s^{H}} &= \frac{1}{-1+\nu}\left( -\frac{156 \nu F^{(1)}_{b c} F^{(1)}_{i a} X_b X_c X_a}{r^4}+\frac{156 F^{(1)}_{b c} F^{(1)}_{i a} X_b X_c
   X_a}{r^4}-\frac{840 \nu F^{(2)}_{i a b c} X_b X_c X_a}{r^4}-\frac{840 F^{(2)}_{i a b c} X_b X_c
   X_a}{r^4}\right.\\
   &\left.+\frac{234 \nu F^{(1)}_{i a} K^{(1)}_{b c} X_b X_c X_a}{r^4}-\frac{126 F^{(1)}_{i a} K^{(1)}_{b c}
   X_b X_c X_a}{r^4}-\frac{306 \nu K^{(1)}_{b c} K^{(1)}_{i a} X_b X_c X_a}{r^4}-\frac{126 K^{(1)}_{b c}
   K^{(1)}_{i a} X_b X_c X_a}{r^4}\right.\\
   &\left.+\frac{126 \nu F^{(1)}_{a b} K^{(1)}_{i c} X_b X_c X_a}{r^4}+\frac{90
   F^{(1)}_{a b} K^{(1)}_{i c} X_b X_c X_a}{r^4}+\frac{210 \nu K^{(2)}_{i a b c} X_b X_c X_a}{r^4}+\frac{3990
   K^{(2)}_{i a b c} X_b X_c X_a}{r^4}\right.\\
   &\left.+\frac{24 \nu F^{(1)}_{a b} X_b X_i X_a}{r^4}+\frac{24 F^{(1)}_{a b}
   X_b X_i X_a}{r^4}+\frac{24 \nu F^{(2)}_{a b} X_b X_i X_a}{r^4}+\frac{24 F^{(2)}_{a b} X_b X_i
   X_a}{r^4}-\frac{24 \nu K^{(1)}_{a b} X_b X_i X_a}{r^4}\right.\\
   &\left.-\frac{48 K^{(1)}_{a b} X_b X_i X_a}{r^4}-\frac{144
   \nu F^{(1)}_{a c} K^{(1)}_{b c} X_b X_i X_a}{r^4}+\frac{360 F^{(1)}_{a c} K^{(1)}_{b c} X_b X_i
   X_a}{r^4}-\frac{36 \nu K^{(1)}_{a c} K^{(1)}_{b c} X_b X_i X_a}{r^4}\right.\\
   &\left.-\frac{288 K^{(1)}_{a c} K^{(1)}_{b c} X_b
   X_i X_a}{r^4}-\frac{24 \nu K^{(2)}_{a b} X_b X_i X_a}{r^4}-\frac{48 K^{(2)}_{a b} X_b X_i
   X_a}{r^4}+\frac{306 \nu F^{(1)}_{a b} F^{(1)}_{c d} X_b X_c X_d X_i X_a}{r^6}\right.\\
   &\left.-\frac{54 F^{(1)}_{a b}
   F^{(1)}_{c d} X_b X_c X_d X_i X_a}{r^6}+\frac{1260 \nu F^{(2)}_{a b c d} X_b X_c X_d X_i
   X_a}{r^6}+\frac{1260 F^{(2)}_{a b c d} X_b X_c X_d X_i X_a}{r^6}\right.\\
   &\left.-\frac{612 \nu F^{(1)}_{a b} K^{(1)}_{c d}
   X_b X_c X_d X_i X_a}{r^6}-\frac{252 F^{(1)}_{a b} K^{(1)}_{c d} X_b X_c X_d X_i X_a}{r^6}+\frac{432 \nu
   K^{(1)}_{a b} K^{(1)}_{c d} X_b X_c X_d X_i X_a}{r^6}\right.\\
   &\left.+\frac{216 K^{(1)}_{a b} K^{(1)}_{c d} X_b X_c X_d X_i
   X_a}{r^6}-\frac{1260 \nu K^{(2)}_{a b c d} X_b X_c X_d X_i X_a}{r^6}-\frac{5040 K^{(2)}_{a b c d} X_b X_c
   X_d X_i X_a}{r^6}-\frac{12 \nu F^{(1)}_{i a} X_a}{r^2}\right.\\
   &\left.-\frac{12 F^{(1)}_{i a} X_a}{r^2}-\frac{36 \nu
   F^{(1)}_{a b} F^{(1)}_{i b} X_a}{r^2}-\frac{36 F^{(1)}_{a b} F^{(1)}_{i b} X_a}{r^2}-\frac{12 \nu F^{(2)}_{ia} X_a}{r^2}-\frac{12 F^{(2)}_{i a} X_a}{r^2}+\frac{18 \nu F^{(1)}_{i b} K^{(1)}_{a b} X_a}{r^2}\right.\\
   &\left.+\frac{18
   F^{(1)}_{i b} K^{(1)}_{a b} X_a}{r^2}+\frac{6 \nu K^{(1)}_{i a} X_a}{r^2}+\frac{30 K^{(1)}_{i a}
   X_a}{r^2}+\frac{90 \nu F^{(1)}_{a b} K^{(1)}_{i b} X_a}{r^2}-\frac{126 F^{(1)}_{a b} K^{(1)}_{i b}
   X_a}{r^2}-\frac{9 \nu K^{(1)}_{a b} K^{(1)}_{i b} X_a}{r^2}\right.\\
   &\left.+\frac{171 K^{(1)}_{a b} K^{(1)}_{i b}
   X_a}{r^2}+\frac{6 \nu K^{(2)}_{i a} X_a}{r^2}+\frac{30 K^{(2)}_{i a} X_a}{r^2}-\frac{24 \nu F^{(1)}_{a b}
   F^{(1)}_{a b} X_i}{5 r^2}-\frac{24 F^{(1)}_{a b} F^{(1)}_{a b} X_i}{5 r^2}+\frac{36 \nu F^{(1)}_{a b}
   K^{(1)}_{a b} X_i}{r^2}\right.\\
   &\left.-\frac{36 F^{(1)}_{a b} K^{(1)}_{a b} X_i}{r^2}+\frac{9 \nu K^{(1)}_{a b} K^{(1)}_{a b}
   X_i}{r^2}+\frac{9 K^{(1)}_{a b} K^{(1)}_{a b} X_i}{r^2} \right)
    \end{split}
\end{equation}

\begin{equation}
    \begin{split}
        \frac{f_i^{NH}}{G_s^{NH}} &= -\frac{2772 X_a X_b X_c X_d X_i F^{(1)}_{a b} K^{(1)}_{c d}}{r^6}+\frac{954 X_a X_b X_c X_d X_i F^{(1)}_{a b}
   F^{(1)}_{c d}}{r^6}-\frac{3780 X_a X_b X_c X_d X_i F^{(2)}_{a b c d}}{r^6}\\
   &+\frac{2376 X_a X_b X_c X_d X_i
   K^{(1)}_{a b} K^{(1)}_{c d}}{r^6}+\frac{11340 X_a X_b X_c X_d X_i K^{(2)}_{a b c d}}{r^6}+\frac{1026 X_a X_b X_c
   F^{(1)}_{i a} K^{(1)}_{b c}}{r^4}\\
   &+\frac{1422 X_a X_b X_c F^{(1)}_{a b} K^{(1)}_{i c}}{r^4}-\frac{288 X_a X_b X_i F^{(1)}_{a c} K^{(1)}_{b c}}{r^4}-\frac{924 X_a X_b X_c F^{(1)}_{i a} F^{(1)}_{b c}}{r^4}+\frac{2520 X_a X_b X_c
   F^{(2)}_{i a b c}}{r^4}\\
   &-\frac{1818 X_a X_b X_c K^{(1)}_{i a} K^{(1)}_{b c}}{r^4}+\frac{180 X_a X_b X_i
   K^{(1)}_{a c} K^{(1)}_{b c}}{r^4}-\frac{8190 X_a X_b X_c K^{(2)}_{i a b c}}{r^4}-\frac{54 X_a K^{(1)}_{a b}
   F^{(1)}_{i b}}{r^2}\\
   &-\frac{126 X_a F^{(1)}_{a b} K^{(1)}_{i b}}{r^2}+\frac{36 X_i F^{(1)}_{a b}
   K^{(1)}_{a b}}{r^2}-\frac{72 X_a X_b X_i F^{(1)}_{a b}}{r^4}+\frac{108 X_a F^{(1)}_{a b} F^{(1)}_{i b}}{r^2}+\frac{72 X_i F^{(1)}_{a b} F^{(1)}_{a b}}{5 r^2}\\
   &-\frac{72 X_a X_b X_i F^{(2)}_{a b}}{r^4}+\frac{120 X_a
   X_b X_i K^{(1)}_{a b}}{r^4}+\frac{99 X_a K^{(1)}_{a b} K^{(1)}_{i b}}{r^2}-\frac{27 X_i K^{(1)}_{a b}
   K^{(1)}_{a b}}{r^2}+\frac{120 X_a X_b X_i K^{(2)}_{a b}}{r^4}\\
   &+\frac{36 X_a F^{(1)}_{i a}}{r^2}+\frac{36 X_a
   F^{(2)}_{i a}}{r^2}-\frac{66 X_a K^{(1)}_{i a}}{r^2}-\frac{66 X_a K^{(2)}_{ia}}{r^2}
    \end{split}
\end{equation}
\subsection{Bending force}
\begin{equation}
\begin{split}
    \frac{f_{i}}{\kappa} = &\frac{72 F^{(1)}_{la}F^{(1)}_{am}-576 F^{(1)}_{la}K^{(1)}_{am}}{r^4}X_i+\frac{-72 F^{(1)}_{lm}-72 F^{(2)}_{lm}-864 F^{(1)}_{la}F^{(1)}_{am}+3456F^{(1)}_{la}K^{(1)}_{am}}{r^6}X_lX_mX_i\\
    &+\frac{432 F^{(1)}_{il}F^{(1)}_{mp}-216F^{(1)}_{lm}K^{(1)}_{ip}}{r^6}X_lX_mX_p+\frac{1836F^{(1)}_{lm}F^{(1)}_{pq}-37800F^{(2)}_{lmpq}-5832F^{(1)}_{lm}K^{(1)}_{pq}}{r^8}X_lX_mX_pX_qX_i
\end{split}
\end{equation}
\subsection{Surface tension force}
\begin{equation}
    \begin{split}
        \frac{f_i}{\sigma} = &\frac{-12F^{(1)}_{lm}-12F^{(2)}_{lm}+144F^{(1)}_{la}K^{(1)}_{am}}{r^4}X_lX_mX_i+\frac{72F^{(1)}_{il}F^{(1)}_{mp}-36F^{(1)}_{lm}K^{(1)}_{ip}}{r^4}X_lX_mX_p\\   &+\frac{18F^{(1)}_{lm}F^{(1)}_{pq}-1890F^{(2)}_{lmpq}-216F^{(1)}_{lm}K^{(1)}_{pq}}{r^6}X_lX_mX_pX_qX_i
    \end{split}
\end{equation}
\clearpage
\section{Matrices values}\label{appendix:matrix}
Due to the length of the matrix Q, the components of the matrix are denoted by $Q_1$ for the first component and $Q_2$ for the second one.

\subsection{Skalak law}
At first order :
\begin{equation}
    M = \begin{pmatrix}
        -\frac{4 \left(60 B (\lambda +1)+10 \lambda  \Sigma +\lambda +10 \Sigma +2 (\lambda +4) C+4\right)}{Ca (2 \lambda +3)
   (19 \lambda +16)} && \frac{6 (2 (\lambda +4) C-5 \lambda )}{Ca (2 \lambda +3) (19 \lambda +16)}\\
        \frac{4 \left(-12B (3 \lambda +2)-6 \lambda  \Sigma +7 \lambda -4 \Sigma +2 (7 \lambda +8) C+8\right)}{Ca (2 \lambda
   +3) (19 \lambda +16)} && -\frac{2 (47 \lambda +6 (7 \lambda +8) C+48)}{Ca \left(38 \lambda ^2+89 \lambda +48\right)}\end{pmatrix}
\end{equation}
\begin{equation}
    Q = \begin{pmatrix}
        \frac{5}{9+6\lambda}\bm{E}+\bm{\omega}\cdot\bm{F}^{(1)}-\bm{F}^{(1)}\cdot\bm{\omega}\\
        \frac{5}{9+6\lambda}\bm{E}+\bm{\omega}\cdot\bm{K}^{(1)}-\bm{K}^{(1)}\cdot\bm{\omega}
    \end{pmatrix}
\end{equation}
At second order in $Ca$ and second order tensors :
\begin{equation}
    M = \begin{pmatrix}
        -\frac{4 \left(60 B (\lambda +1)+10 \lambda  \Sigma +\lambda +10 \Sigma +2 (\lambda +4) C+4\right)}{Ca (2 \lambda +3)
   (19 \lambda +16)} && \frac{6 (2 (\lambda +4) C-5 \lambda )}{Ca (2 \lambda +3) (19 \lambda +16)}\\
        \frac{4 \left(-12B (3 \lambda +2)-6 \lambda  \Sigma +7 \lambda -4 \Sigma +2 (7 \lambda +8) C+8\right)}{Ca (2 \lambda
   +3) (19 \lambda +16)} && -\frac{2 (47 \lambda +6 (7 \lambda +8) C+48)}{Ca \left(38 \lambda ^2+89 \lambda +48\right)}\end{pmatrix}
\end{equation}
\begin{equation}
    \begin{split}
    &Q_1 = sd_2\left[\bm{F}^{(1)}\cdot\bm{F}^{(1)}\right] \left(\frac{576 B \left(601 \lambda ^3+2507 \lambda ^2+2908 \lambda +984\right)}{7 Ca (2 \lambda +3)^2 (19 \lambda
   +16)^2}+\frac{\left(39456 \lambda ^3+179712 \lambda ^2+213408 \lambda +71424\right) \Sigma }{7 Ca (2 \lambda +3)^2 (19 \lambda
   +16)^2}\right.\\
   &\left.+\frac{-20648 \lambda ^3-100996 \lambda ^2-145664 \lambda -40 \left(166 \lambda ^3+3659 \lambda ^2+7684 \lambda +4416\right)
   C-72192}{7 Ca (2 \lambda +3)^2 (19 \lambda +16)^2}\right)\\
   &+sd_2\left[\bm{F}^{(1)}\cdot\bm{K}^{(1)}\right] \left(\frac{432 B \left(12 \lambda ^3-241
   \lambda ^2-454 \lambda -192\right)}{7 Ca (2 \lambda +3)^2 (19 \lambda +16)^2}+\frac{\left(864 \lambda ^3-17352 \lambda ^2-32688
   \lambda -13824\right) \Sigma }{7 Ca (2 \lambda +3)^2 (19 \lambda +16)^2}\right.\\
   &\left.+\frac{-10632 \lambda ^3+59556 \lambda ^2+131784 \lambda +120
   \left(-158 \lambda ^3+1421 \lambda ^2+3862 \lambda +2400\right) C+81792}{7 Ca (2 \lambda +3)^2 (19 \lambda
   +16)^2}\right)\\
   &+\frac{9 sd_2\left[\bm{K}^{(1)}\cdot\bm{K}^{(1)}\right] \left(-7262 \lambda ^3-19379 \lambda ^2-12756 \lambda +2 \left(2486 \lambda ^3+4567 \lambda ^2+608 \lambda
   -1536\right) C-1728\right)}{7 Ca (2 \lambda +3)^2 (19 \lambda +16)^2}\\
   &+\left(\frac{10 sd_2\left[\bm{E}\cdot\bm{F}^{(1)}\right] \left(76 \lambda ^2-107
   \lambda -144\right)}{7 (2 \lambda +3)^2 (19 \lambda +16)}+\frac{75 sd_2\left[\bm{E}\cdot\bm{K}^{(1)}\right] \left(4 \lambda ^2+19 \lambda +12\right) }{7 (2 \lambda +3)^2 (19
   \lambda +16)}\right)\\
   &+\bm{\omega}\cdot\bm{F}^{(2)}-\bm{F}^{(2)}\cdot\bm{\omega}+\bm{\Omega}\cdot\bm{F}^{(2)}-\bm{F}^{(2)}\cdot\bm{\Omega}
    \end{split}
\end{equation}

\begin{equation}
    \begin{split}
    &Q_2 = sd_2\left[\bm{F}^{(1)}\cdot\bm{F}^{(1)}\right] \left(\frac{576 B \left(239 \lambda ^3+1388 \lambda ^2+1932 \lambda +816\right)}{7 Ca (2 \lambda +3)^2 (19 \lambda
   +16)^2}+\frac{\left(12000 \lambda ^3+100320 \lambda ^2+154560 \lambda +69120\right) \Sigma }{7 Ca (2 \lambda +3)^2 (19 \lambda
   +16)^2}\right.\\
   &\left.+\frac{-6856 \lambda ^3-18612 \lambda ^2+11392 \lambda +8 \left(19946 \lambda ^3+68877 \lambda ^2+83608 \lambda +34944\right)
   C+24576}{7 Ca (2 \lambda +3)^2 (19 \lambda +16)^2}\right)\\
   &+sd_2\left[\bm{F}^{(1)}\cdot\bm{K}^{(1)}\right] \left(-\frac{144 B \left(176 \lambda ^3+1717
   \lambda ^2+2888 \lambda +1344\right)}{7 Ca (2 \lambda +3)^2 (19 \lambda +16)^2}+\frac{\left(-4224 \lambda ^3-41208 \lambda ^2-69312
   \lambda -32256\right) \Sigma }{7 Ca (2 \lambda +3)^2 (19 \lambda +16)^2}\right.\\
   &\left.+\frac{34296 \lambda ^3+178452 \lambda ^2+171408 \lambda -24
   \left(17118 \lambda ^3+59831 \lambda ^2+71684 \lambda +28992\right) C+25344}{7 Ca (2 \lambda +3)^2 (19 \lambda
   +16)^2}\right)\\
   &+\frac{9 sd_2\left[\bm{K}^{(1)}\cdot\bm{K}^{(1)}\right] \left(-18326 \lambda ^3-60607 \lambda ^2-61608 \lambda +2 \left(13758 \lambda ^3+48931 \lambda ^2+57664
   \lambda +22272\right) C-19584\right)}{7 Ca (2 \lambda +3)^2 (19 \lambda +16)^2}\\
   &+ \left(\frac{5 sd_2\left[\bm{E}\cdot\bm{K}^{(1)}\right] \left(112 \lambda ^2+451
   \lambda +312\right)}{7 (2 \lambda +3)^2 (19 \lambda +16)}-\frac{40 sd_2\left[\bm{E}\cdot\bm{F}^{(1)}\right] \left(19 \lambda ^3+35 \lambda ^2+73 \lambda
   +48\right)}{7 (2 \lambda +3)^2 (19 \lambda +16)}\right)\\
   &+\bm{\omega}\cdot\bm{K}^{(2)}-\bm{K}^{(2)}\cdot\bm{\omega}+\bm{\Omega}\cdot\bm{F}^{(2)}-\bm{F}^{(2)}\cdot\bm{\Omega}
    \end{split}
\end{equation}
and for fourth order tensors :
\begin{equation}
    M = \begin{pmatrix}
   -\frac{40 \left(540B (\lambda +1)+27 \lambda  \Sigma +\lambda +27 \Sigma +2 (\lambda +2) C+2\right)}{3 Ca (10 \lambda
   +11) (17 \lambda +16)} && \frac{10 (-35 \lambda +20 (\lambda +2) C-16)}{3 Ca (10 \lambda +11) (17 \lambda +16)}\\
        \frac{8 \left(-180B (5 \lambda +4)-45 \lambda  \Sigma +55 \lambda -36 \Sigma +2 (55 \lambda +56) C+56\right)}{3 Ca (10
   \lambda +11) (17 \lambda +16)} && -\frac{2 (1135 \lambda +20 (55 \lambda +56) C+1136)}{3 Ca \left(170 \lambda ^2+347 \lambda +176\right)}
    \end{pmatrix}
\end{equation}

\begin{equation}
    \begin{split}
    &Q_1 = sd_4\left[\bm{F}^{(1)}\cdot\bm{F}^{(1)}\right] \left(\frac{96 B \left(898 \lambda ^3+3255 \lambda ^2+3701 \lambda +1344\right)}{7 Ca (2 \lambda +3) (10 \lambda +11) (17\lambda +16) (19 \lambda +16)}\right.\\
    &\left.+\frac{\left(-3360 \lambda ^3+49680 \lambda ^2+99120 \lambda +46080\right) \Sigma }{105 Ca (2 \lambda
   +3) (10 \lambda +11) (17 \lambda +16) (19 \lambda +16)}\right.\\
   &\left.+\frac{-44720 \lambda ^3-164136 \lambda ^2-180472 \lambda -8 \left(7190 \lambda
   ^3+33969 \lambda ^2+45418 \lambda +18528\right) C-62592}{105 Ca (2 \lambda +3) (10 \lambda +11) (17 \lambda +16) (19 \lambda
   +16)}\right)\\
   &+sd_4\left[\bm{F}^{(1)}\cdot\bm{K}^{(1)}\right] \left(-\frac{48 B \left(30390 \lambda ^3+102287 \lambda ^2+111464 \lambda +39464\right)}{35 Ca (2
   \lambda +3) (10 \lambda +11) (17 \lambda +16) (19 \lambda +16)}\right.\\
   &\left.+\frac{\left(-154800 \lambda ^3-534648 \lambda ^2-603696 \lambda
   -221376\right) \Sigma }{105 Ca (2 \lambda +3) (10 \lambda +11) (17 \lambda +16) (19 \lambda +16)}\right.\\
   &\left.+\frac{-51060 \lambda ^3-91134
   \lambda ^2-63408 \lambda +240 \left(1076 \lambda ^3+3936 \lambda ^2+4517 \lambda +1664\right) C-18048}{105 Ca (2 \lambda +3)
   (10 \lambda +11) (17 \lambda +16) (19 \lambda +16)}\right)\\
   &-\frac{2 sd_4\left[\bm{K}^{(1)}\cdot\bm{K}^{(1)}\right] \left(4530 \lambda ^3+953 \lambda ^2-14504 \lambda +6
   \left(7070 \lambda ^3+21963 \lambda ^2+21896 \lambda +7136\right) C-9984\right)}{35 Ca (2 \lambda +3) (10 \lambda +11) (17
   \lambda +16) (19 \lambda +16)}\\
   &+ \left(\frac{3 sd_4\left[\bm{E}\cdot\bm{F}^{(1)}\right] \left(170 \lambda ^2+347 \lambda +176\right)}{7 (2 \lambda +3) (10 \lambda +11) (17
   \lambda +16)}-\frac{\left(410 sd_4\left[\bm{E}\cdot\bm{K}^{(1)}\right] \lambda ^2+1061 \lambda +608\right)}{21 (2 \lambda +3) (10 \lambda +11) (17 \lambda +16)}\right)\\
   &+\bm{\omega}\cdot\bm{F}^{(2)}-\bm{F}^{(2)}\cdot\bm{\omega}
    \end{split}
\end{equation}

\begin{equation}
    \begin{split}
    &Q_2 = sd_4\left[\bm{F}^{(1)}\cdot\bm{F}^{(1)}\right] \left(-\frac{48 B \left(614 \lambda ^3+561 \lambda ^2-620 \lambda -576\right)}{7 Ca (2 \lambda +3) (10 \lambda +11) (17
   \lambda +16) (19 \lambda +16)}\right.\\
   &\left.-\frac{\left(146640 \lambda ^3+296568 \lambda ^2+154464 \lambda +4608\right) \Sigma }{105 Ca (2 \lambda
   +3) (10 \lambda +11) (17 \lambda +16) (19 \lambda +16)}\right.\\
   &\left.-\frac{-241000 \lambda ^3-788316 \lambda ^2-878624 \lambda -64 \left(16580 \lambda
   ^3+54948 \lambda ^2+60247 \lambda +21840\right) C-330240}{105 Ca (2 \lambda +3) (10 \lambda +11) (17 \lambda +16) (19 \lambda
   +16)}\right)\\
   &+sd_4\left[\bm{F}^{(1)}\cdot\bm{K}^{(1)}\right] \left(-\frac{48 B \left(9110 \lambda ^3+28639 \lambda ^2+29720 \lambda +10336\right)}{35 Ca (2
   \lambda +3) (10 \lambda +11) (17 \lambda +16) (19 \lambda +16)}\right.\\
   &\left.-\frac{\left(27120 \lambda ^3+85560 \lambda ^2+112512 \lambda +54528\right)
   \Sigma }{105 Ca (2 \lambda +3) (10 \lambda +11) (17 \lambda +16) (19 \lambda +16)}\right.\\
   &\left.-\frac{1059120 \lambda ^3+3408480 \lambda ^2+3717696
   \lambda +60 \left(33846 \lambda ^3+116421 \lambda ^2+132352 \lambda +49664\right) C+1365504}{105 Ca (2 \lambda +3) (10 \lambda
   +11) (17 \lambda +16) (19 \lambda +16)}\right)\\
   &+\frac{2 sd_4\left[\bm{K}^{(1)}\cdot\bm{K}^{(1)}\right] \left(1550 \lambda ^3+51575 \lambda ^2+111008 \lambda +6 \left(17250
   \lambda ^3+67749 \lambda ^2+86080 \lambda +35456\right) C+60672\right)}{35 Ca (2 \lambda +3) (10 \lambda +11) (17 \lambda +16)
   (19 \lambda +16)}\\
   &+ \left(-\frac{sd_4\left[\bm{E}\cdot\bm{F}^{(1)}\right] \left(340 \lambda ^3-156 \lambda ^2-1383 \lambda -880\right)}{7 (2 \lambda +3) (10 \lambda +11)
   (17 \lambda +16)}-\frac{sd_4\left[\bm{E}\cdot\bm{K}^{(1)}\right]\left(250 \lambda ^2+1045 \lambda +784\right)}{21 (2 \lambda +3) (10 \lambda +11) (17 \lambda +16)}\right)\\
   &+\bm{\omega}\cdot\bm{K}^{(2)}-\bm{K}^{(2)}\cdot\bm{\omega}
    \end{split}
\end{equation}

\subsection{Hooke law}
At first order :
\begin{equation}
    M = \begin{pmatrix}
        \frac{4 (B (10 \lambda  \Sigma +\lambda +\nu (-10 \lambda  \Sigma +\lambda -10 \Sigma +4)+10 \Sigma +4)-60 (\lambda +1) (\nu-1))}{B
   Ca (2 \lambda +3) (19 \lambda +16) (\nu-1)} && -\frac{6 ((7 \lambda +8) \nu-5 \lambda )}{Ca (2 \lambda +3) (19 \lambda +16) (\nu-1)}\\
  -\frac{4 (B (-6 \lambda  \Sigma +7 \lambda +\nu (6 \lambda  \Sigma +7 \lambda +4 \Sigma +8)-4 \Sigma +8)+12 (3 \lambda +2)
   (\nu-1))}{B Ca (2 \lambda +3) (19 \lambda +16) (\nu-1)}&&\frac{2 (\lambda  (47-5 \nu)+48)}{Ca (2 \lambda +3) (19 \lambda +16) (\nu-1)}
    \end{pmatrix}
\end{equation}
\begin{equation}
    Q = \begin{pmatrix}
        \frac{5}{9+6\lambda}\bm{E}+\bm{\omega}\cdot\bm{F}^{(1)}-\bm{F}^{(1)}\cdot\bm{\omega}\\
        \frac{5}{9+6\lambda}\bm{E}+\bm{\omega}\cdot\bm{K}^{(1)}-\bm{K}^{(1)}\cdot\bm{\omega}
    \end{pmatrix}
\end{equation}
At second order in $Ca$ and second order tensors :
\begin{equation}
    M = \begin{pmatrix}
        \frac{4 (B (10 \lambda  \Sigma +\lambda +\nu (-10 \lambda  \Sigma +\lambda -10 \Sigma +4)+10 \Sigma +4)-60 (\lambda +1) (\nu-1))}{B
   Ca (2 \lambda +3) (19 \lambda +16) (\nu-1)} && -\frac{6 ((7 \lambda +8) \nu-5 \lambda )}{Ca (2 \lambda +3) (19 \lambda +16) (\nu-1)}\\
  -\frac{4 (B (-6 \lambda  \Sigma +7 \lambda +\nu (6 \lambda  \Sigma +7 \lambda +4 \Sigma +8)-4 \Sigma +8)+12 (3 \lambda +2)
   (\nu-1))}{B Ca (2 \lambda +3) (19 \lambda +16) (\nu-1)}&&\frac{2 (\lambda  (47-5 \nu)+48)}{Ca (2 \lambda +3) (19 \lambda +16) (\nu-1)}
    \end{pmatrix}
\end{equation}

\begin{equation}
    \begin{split}
    &Q_1 = sd_2\left[\bm{F}^{(1)}\cdot\bm{F}^{(1)}\right] \left(\frac{576 B \left(601 \lambda ^3+2507 \lambda ^2+2908 \lambda +984\right)}{7 Ca (2 \lambda +3)^2 (19 \lambda
   +16)^2}\right.\\
   &\left.+\frac{\Sigma  \left(-39456 \lambda ^3-179712 \lambda ^2-213408 \lambda +39456 \lambda ^3 \nu+179712 \lambda ^2 \nu+213408
   \lambda  \nu+71424 \nu-71424\right)}{7 Ca (2 \lambda +3)^2 (19 \lambda +16)^2 (\nu-1)}\right.\\
   &\left.+\frac{20648 \lambda ^3+100996
   \lambda ^2+145664 \lambda -26168 \lambda ^3 \nu-31756 \lambda ^2 \nu+32416 \lambda  \nu+43008 \nu+72192}{7 Ca
   (2 \lambda +3)^2 (19 \lambda +16)^2 (\nu-1)}\right)\\
   &+sd_2\left[\bm{F}^{(1)}\cdot\bm{K}^{(1)}\right] \left(\frac{432 B \left(12 \lambda ^3-241 \lambda ^2-454
   \lambda -192\right)}{7 Ca (2 \lambda +3)^2 (19 \lambda +16)^2}\right.\\
   &\left.+\frac{\Sigma  \left(-864 \lambda ^3+17352 \lambda ^2+32688 \lambda +864
   \lambda ^3 \nu-17352 \lambda ^2 \nu-32688 \lambda  \nu-13824 \nu+13824\right)}{7 Ca (2 \lambda +3)^2 (19
   \lambda +16)^2 (\nu-1)}\right.\\
   &\left.+\frac{10632 \lambda ^3-59556 \lambda ^2-131784 \lambda -20856 \lambda ^3 \nu-77172 \lambda ^2
   \nu-129288 \lambda  \nu-77184 \nu-81792}{7 Ca (2 \lambda +3)^2 (19 \lambda +16)^2 (\nu-1)}\right)\\
   &-\frac{9
   sd_2\left[\bm{K}^{(1)}\cdot\bm{K}^{(1)}\right] \left(-7262 \lambda ^3-19379 \lambda ^2-12756 \lambda +5 \left(866 \lambda ^3+2973 \lambda ^2+3076 \lambda +960\right)
   \nu-1728\right)}{7 Ca (2 \lambda +3)^2 (19 \lambda +16)^2 (\nu-1)}\\
   &+ \left(\frac{10 sd_2\left[\bm{E}\cdot\bm{F}^{(1)}\right] \left(76 \lambda ^2-107
   \lambda -144\right)}{7 (2 \lambda +3)^2 (19 \lambda +16)}+\frac{75 sd_2\left[\bm{E}\cdot\bm{K}^{(1)}\right] \left(4 \lambda ^2+19 \lambda +12\right) }{7 (2 \lambda +3)^2 (19
   \lambda +16)}\right)\\
   &+\bm{\omega}\cdot\bm{F}^{(2)}-\bm{F}^{(2)}\cdot\bm{\omega}+\bm{\Omega}\cdot\bm{F}^{(2)}-\bm{F}^{(2)}\cdot\bm{\Omega}
    \end{split}
\end{equation}

\begin{equation}
    \begin{split}
    &Q_2 = sd_2\left[\bm{F}^{(1)}\cdot\bm{F}^{(1)}\right] \left(\frac{576 B \left(239 \lambda ^3+1388 \lambda ^2+1932 \lambda +816\right)}{7 Ca (2 \lambda +3)^2 (19 \lambda
   +16)^2}\right.\\
   &\left.-\frac{\Sigma  \left(12000 \lambda ^3+100320 \lambda ^2+154560 \lambda -12000 \lambda ^3 \nu-100320 \lambda ^2 \nu-154560
   \lambda  \nu-69120 \nu+69120\right)}{7 Ca (2 \lambda +3)^2 (19 \lambda +16)^2 (\nu-1)}\right.\\
   &\left.-\frac{-6856 \lambda ^3-18612
   \lambda ^2+11392 \lambda +81304 \lambda ^3 \nu+272988 \lambda ^2 \nu+322112 \lambda  \nu+132096 \nu+24576}{7
   Ca (2 \lambda +3)^2 (19 \lambda +16)^2 (\nu-1)}\right)\\
   &+sd_2\left[\bm{F}^{(1)}\cdot\bm{K}^{(1)}\right] \left(-\frac{144 B \left(176 \lambda ^3+1717 \lambda
   ^2+2888 \lambda +1344\right)}{7 Ca (2 \lambda +3)^2 (19 \lambda +16)^2}\right.\\
   &\left.-\frac{\Sigma  \left(-4224 \lambda ^3-41208 \lambda ^2-69312
   \lambda +4224 \lambda ^3 \nu+41208 \lambda ^2 \nu+69312 \lambda  \nu+32256 \nu-32256\right)}{7 Ca (2 \lambda
   +3)^2 (19 \lambda +16)^2 (\nu-1)}\right.\\
   &\left.-\frac{34296 \lambda ^3+178452 \lambda ^2+171408 \lambda -94920 \lambda ^3 \nu-414780 \lambda ^2
   \nu-560880 \lambda  \nu-241920 \nu+25344}{7 Ca (2 \lambda +3)^2 (19 \lambda +16)^2 (\nu-1)}\right)\\
   &-\frac{9
   sd_2\left[\bm{K}^{(1)}\cdot\bm{K}^{(1)}\right] \left(-18326 \lambda ^3-60607 \lambda ^2-61608 \lambda +\left(8754 \lambda ^3+32693 \lambda ^2+38952 \lambda +14976\right)
   \nu-19584\right)}{7 Ca (2 \lambda +3)^2 (19 \lambda +16)^2 (\nu-1)}\\
   &+ \left(\frac{5sd_2\left[\bm{E}\cdot\bm{K}^{(1)}\right] \left(112 \lambda ^2+451 \lambda
   +312\right)}{7 (2 \lambda +3)^2 (19 \lambda +16)}-\frac{40 sd_2\left[\bm{E}\cdot\bm{F}^{(1)}\right] \left(19 \lambda ^3+35 \lambda ^2+73 \lambda +48\right)}{7 (2
   \lambda +3)^2 (19 \lambda +16)}\right)\\
   &+\bm{\omega}\cdot\bm{K}^{(2)}-\bm{K}^{(2)}\cdot\bm{\omega}+\bm{\Omega}\cdot\bm{F}^{(2)}-\bm{F}^{(2)}\cdot\bm{\Omega}
    \end{split}
\end{equation}
and for fourth order tensors :
\begin{equation}
    M = \begin{pmatrix}
       -\frac{40 (B (-27 \lambda  \Sigma -\lambda +\nu (27 \lambda  \Sigma -\lambda +27 \Sigma -2)-27 \Sigma -2)+540 (\lambda +1)
   (\nu-1))}{3 B Ca (10 \lambda +11) (17 \lambda +16) (\nu-1)} && -\frac{10 (-35 \lambda +(55 \lambda +56) \nu-16)}{3 Ca (10 \lambda +11) (17 \lambda +16) (\nu-1)}\\
        -\frac{8 (B (-45 \lambda  \Sigma +55 \lambda +\nu (45 \lambda  \Sigma +55 \lambda +36 \Sigma +56)-36 \Sigma +56)+180 (5 \lambda +4)
   (\nu-1))}{3 B Ca (10 \lambda +11) (17 \lambda +16) (\nu-1)} && \frac{2 (1135 \lambda -(35 \lambda +16) \nu+1136)}{3 Ca (10 \lambda +11) (17 \lambda +16) (\nu-1)}
    \end{pmatrix}
\end{equation}

\begin{equation}
    \begin{split}
    &Q_1 = sd_4\left[\bm{F}^{(1)}\cdot\bm{F}^{(1)}\right] \left(\frac{48 B \left(2176 \lambda ^3+7818 \lambda ^2+8861 \lambda +3216\right)}{7 Ca (2 \lambda +3) (10 \lambda +11) (17
   \lambda +16) (19 \lambda +16)}\right.\\
   &\left.+\frac{\Sigma  \left(3360 \lambda ^3-49680 \lambda ^2-99120 \lambda -3360 \lambda ^3 \nu+49680 \lambda ^2
   \nu+99120 \lambda  \nu+46080 \nu-46080\right)}{105 Ca (2 \lambda +3) (10 \lambda +11) (17 \lambda +16) (19 \lambda
   +16) (\nu-1)}\right.\\
   &\left.+\frac{44720 \lambda ^3+164136 \lambda ^2+180472 \lambda -17600 \lambda ^3 \nu-24384 \lambda ^2 \nu+2072
   \lambda  \nu+8832 \nu+62592}{105 Ca (2 \lambda +3) (10 \lambda +11) (17 \lambda +16) (19 \lambda +16)
   (\nu-1)}\right)\\
   &+sd_4\left[\bm{F}^{(1)}\cdot\bm{K}^{(1)}\right] \left(-\frac{24 B \left(62680 \lambda ^3+211114 \lambda ^2+230223 \lambda +81568\right)}{35
   Ca (2 \lambda +3) (10 \lambda +11) (17 \lambda +16) (19 \lambda +16)}\right.\\
   &\left.+\frac{\Sigma  \left(154800 \lambda ^3+534648 \lambda ^2+603696
   \lambda -154800 \lambda ^3 \nu-534648 \lambda ^2 \nu-603696 \lambda  \nu-221376 \nu+221376\right)}{105 Ca (2
   \lambda +3) (10 \lambda +11) (17 \lambda +16) (19 \lambda +16) (\nu-1)}\right.\\
   &\left.+\frac{51060 \lambda ^3+91134 \lambda ^2+63408 \lambda -149700
   \lambda ^3 \nu-424854 \lambda ^2 \nu-390528 \lambda  \nu-117888 \nu+18048}{105 Ca (2 \lambda +3) (10 \lambda
   +11) (17 \lambda +16) (19 \lambda +16) (\nu-1)}\right)\\
   &+\frac{2 sd_4\left[\bm{K}^{(1)}\cdot\bm{K}^{(1)}\right] \left(4530 \lambda ^3+953 \lambda ^2-14504 \lambda +5
   \left(1878 \lambda ^3+5519 \lambda ^2+4888 \lambda +1344\right) \nu-9984\right)}{35 Ca (2 \lambda +3) (10 \lambda +11) (17
   \lambda +16) (19 \lambda +16) (\nu-1)}\\
   &+ \left(\frac{3 sd_4\left[\bm{E}\cdot\bm{F}^{(1)}\right] \left(170 \lambda ^2+347 \lambda +176\right)}{7 (2 \lambda +3) (10
   \lambda +11) (17 \lambda +16)}-\frac{sd_4\left[\bm{E}\cdot\bm{K}^{(1)}\right]\left(410 \lambda ^2+1061 \lambda +608\right) }{21 (2 \lambda +3) (10 \lambda +11) (17 \lambda
   +16)}\right)\\
   &+\bm{\omega}\cdot\bm{F}^{(2)}-\bm{F}^{(2)}\cdot\bm{\omega}
    \end{split}
\end{equation}

\begin{equation}
    \begin{split}
    &Q_2 = sd_4\left[\bm{F}^{(1)}\cdot\bm{F}^{(1)}\right] \left(-\frac{144 B \left(1530 \lambda ^3+2679 \lambda ^2+912 \lambda -256\right)}{35 Ca (2 \lambda +3) (10 \lambda +11) (17
   \lambda +16) (19 \lambda +16)}\right.\\
   &\left.-\frac{\Sigma  \left(-146640 \lambda ^3-296568 \lambda ^2-154464 \lambda +146640 \lambda ^3 \nu+296568
   \lambda ^2 \nu+154464 \lambda  \nu+4608 \nu-4608\right)}{105 Ca (2 \lambda +3) (10 \lambda +11) (17 \lambda +16) (19
   \lambda +16) (\nu-1)}\right.\\
   &\left.-\frac{241000 \lambda ^3+788316 \lambda ^2+878624 \lambda +485720 \lambda ^3 \nu+1604676 \lambda ^2
   \nu+1757344 \lambda  \nu+637440 \nu+330240}{105 Ca (2 \lambda +3) (10 \lambda +11) (17 \lambda +16) (19 \lambda +16)
   (\nu-1)}\right)\\
   &+sd_4\left[\bm{F}^{(1)}\cdot\bm{K}^{(1)}\right] \left(-\frac{48 B \left(8350 \lambda ^3+26023 \lambda ^2+26802 \lambda +9280\right)}{35 Ca
   (2 \lambda +3) (10 \lambda +11) (17 \lambda +16) (19 \lambda +16)}\right.\\
   &\left.-\frac{\Sigma  \left(-27120 \lambda ^3-85560 \lambda ^2-112512 \lambda
   +27120 \lambda ^3 \nu+85560 \lambda ^2 \nu+112512 \lambda  \nu+54528 \nu-54528\right)}{105 Ca (2 \lambda +3)
   (10 \lambda +11) (17 \lambda +16) (19 \lambda +16) (\nu-1)}\right.\\
   &\left.-\frac{-1059120 \lambda ^3-3408480 \lambda ^2-3717696 \lambda -146280
   \lambda ^3 \nu-797364 \lambda ^2 \nu-1198656 \lambda  \nu-545280 \nu-1365504}{105 Ca (2 \lambda +3) (10
   \lambda +11) (17 \lambda +16) (19 \lambda +16) (\nu-1)}\right)\\
   &-\frac{2 sd_4\left[\bm{K}^{(1)}\cdot\bm{K}^{(1)}\right] \left(1550 \lambda ^3+51575 \lambda ^2+111008
   \lambda +5 \left(4202 \lambda ^3+16289 \lambda ^2+21344 \lambda +9216\right) \nu+60672\right)}{35 Ca (2 \lambda +3) (10 \lambda
   +11) (17 \lambda +16) (19 \lambda +16) (\nu-1)}\\
   &+ \left(-\frac{sd_4\left[\bm{E}\cdot\bm{F}^{(1)}\right] \left(340 \lambda ^3-156 \lambda ^2-1383 \lambda
   -880\right)}{7 (2 \lambda +3) (10 \lambda +11) (17 \lambda +16)}-\frac{sd_4\left[\bm{E}\cdot\bm{K}^{(1)}\right]\left(250 \lambda ^2+1045 \lambda +784\right)}{21 (2 \lambda
   +3) (10 \lambda +11) (17 \lambda +16)}\right)\\
   &+\bm{\omega}\cdot\bm{K}^{(2)}-\bm{K}^{(2)}\cdot\bm{\omega}
    \end{split}
\end{equation}

\subsection{Neo-Hookean law}
At first order :
\begin{equation}
    M = \begin{pmatrix}
        -\frac{4 \left(60B (\lambda +1)+10 \lambda  \Sigma +3 \lambda +10 \Sigma +12\right)}{Ca (2 \lambda +3) (19 \lambda +16)}&&\frac{6 (8-3 \lambda )}{Ca \left(38 \lambda ^2+89 \lambda +48\right)}\\
     -\frac{4 \left(36B \lambda +24+3 \lambda  (2 \Sigma -7)+4 (\Sigma -6)\right)}{Ca (2 \lambda +3) (19 \lambda +16)}&&-\frac{2 (89 \lambda +96)}{Ca \left(38 \lambda ^2+89 \lambda +48\right)}
    \end{pmatrix}
\end{equation}
\begin{equation}
    Q = \begin{pmatrix}
        \frac{5}{9+6\lambda}\bm{E}+\bm{\omega}\cdot\bm{F}^{(1)}-\bm{F}^{(1)}\cdot\bm{\omega}\\
        \frac{5}{9+6\lambda}\bm{E}+\bm{\omega}\cdot\bm{K}^{(1)}-\bm{K}^{(1)}\cdot\bm{\omega}
    \end{pmatrix}
\end{equation}
At second order in $Ca$ and second order tensors :
\begin{equation}
    M = \begin{pmatrix}
        -\frac{4 \left(60B (\lambda +1)+10 \lambda  \Sigma +3 \lambda +10 \Sigma +12\right)}{Ca (2 \lambda +3) (19 \lambda +16)}&&\frac{6 (8-3 \lambda )}{Ca \left(38 \lambda ^2+89 \lambda +48\right)}\\
     -\frac{4 \left(36B \lambda +24+3 \lambda  (2 \Sigma -7)+4 (\Sigma -6)\right)}{Ca (2 \lambda +3) (19 \lambda +16)}&&-\frac{2 (89 \lambda +96)}{Ca \left(38 \lambda ^2+89 \lambda +48\right)}
    \end{pmatrix}
\end{equation}

\begin{equation}
    \begin{split}
    &Q_1 = sd_2\left[\bm{F}^{(1)}\cdot\bm{F}^{(1)}\right] \left(\frac{576 B \left(601 \lambda ^3+2507 \lambda ^2+2908 \lambda +984\right)}{7 Ca (2 \lambda +3)^2 (19 \lambda
   +16)^2}\right.\\
   &\left.+\frac{\left(39456 \lambda ^3+179712 \lambda ^2+213408 \lambda +71424\right) \Sigma }{7 Ca (2 \lambda +3)^2 (19 \lambda
   +16)^2}\right.\\
   &\left.+\frac{4328 \lambda ^3-46844 \lambda ^2-116896 \lambda -89088}{7 Ca (2 \lambda +3)^2 (19 \lambda +16)^2}\right)\\
   &+sd_2\left[\bm{F}^{(1)}\cdot\bm{K}^{(1)}\right] \left(\frac{432 B \left(12 \lambda ^3-241 \lambda ^2-454 \lambda -192\right)}{7 Ca (2 \lambda +3)^2 (19 \lambda
   +16)^2}\right.\\
   &\left.+\frac{\left(864 \lambda ^3-17352 \lambda ^2-32688 \lambda -13824\right) \Sigma }{7 Ca (2 \lambda +3)^2 (19 \lambda
   +16)^2}+\frac{-190104 \lambda ^3-569028 \lambda ^2-598632 \lambda -164736}{7 Ca (2 \lambda +3)^2 (19 \lambda +16)^2}\right)\\
   &+\frac{9 sd_2\left[\bm{K}^{(1)}\cdot\bm{K}^{(1)}\right]
   \left(16558 \lambda ^3+72811 \lambda ^2+103404 \lambda +44352\right)}{7 Ca (2 \lambda +3)^2 (19 \lambda +16)^2}\\
   &+
   \left(\frac{10 sd_2\left[\bm{E}\cdot\bm{F}^{(1)}\right] \left(76 \lambda ^2-107 \lambda -144\right)}{7 (2 \lambda +3)^2 (19 \lambda +16)}+\frac{75 sd_2\left[\bm{E}\cdot\bm{K}^{(1)}\right] \left(4 \lambda ^2+19
   \lambda +12\right)}{7 (2 \lambda +3)^2 (19 \lambda +16)}\right)\\
   &+\bm{\omega}\cdot\bm{F}^{(2)}-\bm{F}^{(2)}\cdot\bm{\omega}+\bm{\Omega}\cdot\bm{F}^{(2)}-\bm{F}^{(2)}\cdot\bm{\Omega}
    \end{split}
\end{equation}

\begin{equation}
    \begin{split}
    &Q_2 = sd_2\left[\bm{F}^{(1)}\cdot\bm{F}^{(1)}\right] \left(\frac{576 B \left(239 \lambda ^3+1388 \lambda ^2+1932 \lambda +816\right)}{7 Ca (2 \lambda +3)^2 (19 \lambda
   +16)^2}\right.\\
   &\left.-\frac{\left(-12000 \lambda ^3-100320 \lambda ^2-154560 \lambda -69120\right) \Sigma }{7 Ca (2 \lambda +3)^2 (19 \lambda
   +16)^2}-\frac{68600 \lambda ^3+238860 \lambda ^2+191680 \lambda +15360}{7 Ca (2 \lambda +3)^2 (19 \lambda +16)^2}\right)\\
   &+sd_2\left[\bm{F}^{(1)}\cdot\bm{K}^{(1)}\right] \left(-\frac{144 B \left(176 \lambda ^3+1717 \lambda ^2+2888 \lambda +1344\right)}{7 Ca (2 \lambda +3)^2 (19 \lambda
   +16)^2}-\frac{\left(4224 \lambda ^3+41208 \lambda ^2+69312 \lambda +32256\right) \Sigma }{7 Ca (2 \lambda +3)^2 (19 \lambda
   +16)^2}\right.\\
   &\left.-\frac{-192552 \lambda ^3-746604 \lambda ^2-741936 \lambda -177408}{7 Ca (2 \lambda +3)^2 (19 \lambda +16)^2}\right)\\
   &-\frac{45 sd_2\left[\bm{K}^{(1)}\cdot\bm{K}^{(1)}\right]
   \left(6066 \lambda ^3+20789 \lambda ^2+22024 \lambda +7296\right)}{7 Ca (2 \lambda +3)^2 (19 \lambda +16)^2}\\
   &+
   \left(\frac{5 sd_2\left[\bm{E}\cdot\bm{K}^{(1)}\right] \left(112 \lambda ^2+451 \lambda +312\right)}{7 (2 \lambda +3)^2 (19 \lambda +16)}-\frac{40 sd_2\left[\bm{E}\cdot\bm{F}^{(1)}\right] \left(19
   \lambda ^3+35 \lambda ^2+73 \lambda +48\right)}{7 (2 \lambda +3)^2 (19 \lambda +16)}\right)\\
   &+\bm{\omega}\cdot\bm{K}^{(2)}-\bm{K}^{(2)}\cdot\bm{\omega}+\bm{\Omega}\cdot\bm{F}^{(2)}-\bm{F}^{(2)}\cdot\bm{\Omega}
    \end{split}
\end{equation}
and for fourth order tensors :
\begin{equation}
    M = \begin{pmatrix}
       -\frac{40 \left(180B (\lambda +1)+9 \lambda  \Sigma +\lambda +9 \Sigma +2\right)}{Ca (10 \lambda +11) (17 \lambda +16)}&&\frac{80-50 \lambda }{170 Ca \lambda ^2+347 Ca \lambda +176 Ca}\\
      -\frac{8 \left(60B (5 \lambda +4)+5 \lambda  (3 \Sigma -11)+12 \Sigma -56\right)}{Ca (10 \lambda +11) (17 \lambda +16)}&&-\frac{2 (745 \lambda +752)}{Ca (10 \lambda +11) (17 \lambda +16)}
    \end{pmatrix}
\end{equation}

\begin{equation}
    \begin{split}
    &Q_1 = sd_4\left[\bm{F}^{(1)}\cdot\bm{F}^{(1)}\right] \left(\frac{96 B \left(898 \lambda ^3+3255 \lambda ^2+3701 \lambda +1344\right)}{7 Ca (2 \lambda +3) (10 \lambda +11) (17
   \lambda +16) (19 \lambda +16)}\right.\\
   &\left.+\frac{\left(-3360 \lambda ^3+49680 \lambda ^2+99120 \lambda +46080\right) \Sigma }{105 Ca (2 \lambda
   +3) (10 \lambda +11) (17 \lambda +16) (19 \lambda +16)}+\frac{-23200 \lambda ^3-92688 \lambda ^2-73736 \lambda -11136}{105 Ca (2
   \lambda +3) (10 \lambda +11) (17 \lambda +16) (19 \lambda +16)}\right)\\
   &+sd_4\left[\bm{F}^{(1)}\cdot\bm{K}^{(1)}\right] \left(-\frac{48 B \left(30390 \lambda ^3+102287
   \lambda ^2+111464 \lambda +39464\right)}{35 Ca (2 \lambda +3) (10 \lambda +11) (17 \lambda +16) (19 \lambda +16)}\right.\\
   &\left.+\frac{\left(-154800
   \lambda ^3-534648 \lambda ^2-603696 \lambda -221376\right) \Sigma }{105 Ca (2 \lambda +3) (10 \lambda +11) (17 \lambda +16) (19
   \lambda +16)}+\frac{38460 \lambda ^3+38826 \lambda ^2-175008 \lambda -148608}{105 Ca (2 \lambda +3) (10 \lambda +11) (17 \lambda +16)
   (19 \lambda +16)}\right)\\
   &+\frac{2 sd_4\left[\bm{K}^{(1)}\cdot\bm{K}^{(1)}\right] \left(-34410 \lambda ^3-57761 \lambda ^2+5768 \lambda +24768\right)}{35 Ca (2 \lambda +3)
   (10 \lambda +11) (17 \lambda +16) (19 \lambda +16)}\\
   &+ \left(\frac{3 sd_4\left[\bm{E}\cdot\bm{F}^{(1)}\right] \left(170 \lambda ^2+347 \lambda +176\right)}{7 (2 \lambda +3)
   (10 \lambda +11) (17 \lambda +16)}-\frac{sd_4\left[\bm{E}\cdot\bm{K}^{(1)}\right]\left(410 \lambda ^2+1061 \lambda +608\right)}{21 (2 \lambda +3) (10 \lambda +11) (17
   \lambda +16)}\right)\\
   &+\bm{\omega}\cdot\bm{F}^{(2)}-\bm{F}^{(2)}\cdot\bm{\omega}
    \end{split}
\end{equation}

\begin{equation}
    \begin{split}
    &Q_2 = sd_4\left[\bm{F}^{(1)}\cdot\bm{F}^{(1)}\right] \left(-\frac{48 B \left(614 \lambda ^3+561 \lambda ^2-620 \lambda -576\right)}{7 Ca (2 \lambda +3) (10 \lambda +11) (17
   \lambda +16) (19 \lambda +16)}\right.\\
   &\left.-\frac{\left(146640 \lambda ^3+296568 \lambda ^2+154464 \lambda +4608\right) \Sigma }{105 Ca (2 \lambda
   +3) (10 \lambda +11) (17 \lambda +16) (19 \lambda +16)}-\frac{-432680 \lambda ^3-1383420 \lambda ^2-1562848 \lambda -609792}{105 Ca (2
   \lambda +3) (10 \lambda +11) (17 \lambda +16) (19 \lambda +16)}\right)\\
   &+sd_4\left[\bm{F}^{(1)}\cdot\bm{K}^{(1)}\right] \left(-\frac{48 B \left(9110 \lambda ^3+28639
   \lambda ^2+29720 \lambda +10336\right)}{35 Ca (2 \lambda +3) (10 \lambda +11) (17 \lambda +16) (19 \lambda +16)}\right.\\
   &\left.-\frac{\left(27120
   \lambda ^3+85560 \lambda ^2+112512 \lambda +54528\right) \Sigma }{105 Ca (2 \lambda +3) (10 \lambda +11) (17 \lambda +16) (19 \lambda
   +16)}-\frac{303720 \lambda ^3+1037412 \lambda ^2+1509312 \lambda +769536}{105 Ca (2 \lambda +3) (10 \lambda +11) (17 \lambda +16) (19
   \lambda +16)}\right)\\
   &-\frac{2 sd_4\left[\bm{K}^{(1)}\cdot\bm{K}^{(1)}\right] \left(265450 \lambda ^3+785425 \lambda ^2+720544 \lambda +201216\right)}{35 Ca (2 \lambda +3)
   (10 \lambda +11) (17 \lambda +16) (19 \lambda +16)}\\
   &+ \left(-\frac{sd_4\left[\bm{E}\cdot\bm{F}^{(1)}\right] \left(340 \lambda ^3-156 \lambda ^2-1383 \lambda -880\right)}{7
   (2 \lambda +3) (10 \lambda +11) (17 \lambda +16)}-\frac{sd_4\left[\bm{E}\cdot\bm{K}^{(1)}\right]\left(250 \lambda ^2+1045 \lambda +784\right)}{21 (2 \lambda +3) (10
   \lambda +11) (17 \lambda +16)}\right)\\
   &+\bm{\omega}\cdot\bm{K}^{(2)}-\bm{K}^{(2)}\cdot\bm{\omega}
    \end{split}
\end{equation}
\end{document}